\documentclass[aps,prb,amsmath,amssymb,floatfix,showpacs,twocolumn,10pt]{revtex4-1}

\usepackage{graphicx}
\usepackage[caption=false]{subfig}
\usepackage{amsmath}
\usepackage{bbold}
\usepackage{latexsym}
\usepackage{color}

\begin{document}


\title{Long-Distance Entanglement of Spin Qubits via Quantum Hall Edge States}



\author{Guang Yang$^1$, Chen-Hsuan Hsu$^1$, Peter Stano$^1$, Jelena Klinovaja$^2$, and Daniel Loss$^{1,2}$}
\affiliation{$^1$RIKEN Center for Emergent Matter Science, Wako, Saitama 351-0198, Japan \\ $^2$Department of Physics, University of Basel, Klingelbergstrasse 82, CH-4056 Basel, Switzerland}



\date{\today}

\begin{abstract}
The implementation of a functional quantum computer involves entangling and coherent manipulation of a large number of qubits. For qubits based on electron spins confined in quantum dots, which are among the most investigated solid-state qubits at present, architectural challenges are often encountered in the design of quantum circuits attempting to assemble the qubits within the very limited space available. Here, we provide a solution to such challenges based on an approach to realizing entanglement of spin qubits over long distances. We show that long-range Ruderman-Kittel-Kasuya-Yosida interaction of confined electron spins can be established by quantum Hall edge states, leading to an exchange coupling of spin qubits. The coupling is anisotropic and can be either Ising-type or XY-type, depending on the spin polarization of the edge state. Such a property, combined with the dependence of the electron spin susceptibility on the chirality of the edge state, can be utilized to gain valuable insights into the topological nature of various quantum Hall states.

\end{abstract}

\pacs{73.43.Fj, 73.63.Kv, 03.67.Lx}


\maketitle


\section{Introduction}

Quantum computers, exploiting entanglement and superposition of quantum mechanical states, promise much better performance than classical computers tackling a collection of important mathematical problems \cite{qcbook}. Over the past few decades, a variety of solid-state systems have been studied for the implementation of qubits, the building blocks of a quantum computer. Among such systems, a very promising candidate \cite{loss98} makes use of the spin of electrons confined in semiconductor quantum dots (QDs). In that scheme, entanglement of qubits is achieved through the direct exchange interaction between confined electrons and manipulation of individual qubits can be realized by magnetic or electrical means. \cite{qcreview} Recent advances in QD technology have established long coherence times \cite{bluhm11} exceeding $0.2$~ms and fast gate-operation times \cite{qcreview} on the order of tens of nanoseconds for spin qubits in QDs. 

With the great progress in the development of quality spin qubits, scalability becomes the next major challenge towards building a functional quantum computer capable of performing fault-tolerant quantum computing \cite{svore05}. The implementation of quantum-error-correction algorithms \cite{childs03} requires that the system reach a size of several thousands of qubits. In practice, however, one faces tremendous difficulties in assembling so many spin qubits, among which entanglement must be selectively established and maintained. Indeed, the nearest-neighbor nature of the direct exchange interaction, the primary source of entanglement, restricts drastically access of each qubit to the rest of the system and thus the space that can be used for installing the quantum circuits. A natural way to overcome such difficulties is to employ means of entangling spin qubits over long distances, which creates extra space for wiring the quantum circuits. In principle, this may be achieved by coupling the spin qubits to an electromagnetic cavity \cite{imamoglu99,childress04,burkard06,trif08}, a floating metallic gate \cite{trifunovic12}, or a dipolar ferromagnet \cite{trifunovic13}. Recently, it was shown that coupling of distant spin qubits can also be realized via photon-assisted cotunneling \cite{stano14}. 
 
In this article, we propose a new mechanism to achieve long-distance entanglement of spin qubits. We make use of the Ruderman-Kittel-Kasuya-Yosida (RKKY) interaction \cite{rkky1,rkky2,rkky3} between confined electron spins in QDs, mediated by the conducting edge states of quantum Hall (QH) liquids \cite{wenbook}, to which the QDs are tunnel coupled \cite{kiyama15}. The spin qubit coupling obtained in such a way is particularly interesting. Depending on whether the edge state is spin-polarized or not, the induced coupling between the spin qubits can be Ising-type and perpendicular to the plane of the system, or XY-type and in-plane. This offers great versatility in the design of large-scale quantum circuits. The advantage of using QH edge states is twofold. First, the edge states and the QDs can be formed in the same material (by top gates) such as a two-dimensional electron gas (2DEG) in GaAs heterostructures. Second, the QH edge states are topologically stable and thus much more robust against disorder effects compared to one-dimensional (1D) conduction channels in nano- or quantum wires.
Moreover, we find that the spin susceptibility of QH edge states manifests the  inequivalence between the opposite directions, ``clockwise'' and ``counterclockwise'', along the QH edge. In chiral edge states, conduction electrons propagate in only one direction, leading to a ``rectified'' spin susceptibility in the propagation direction of electrons. In non-chiral edge states, the spin susceptibility is nonzero in both directions along the QH edge, but with different magnitudes.  The spin susceptibility has the same type of anisotropy as the coupling between qubits. Thus, measuring the spatial dependence of the spin susceptibility \cite{stano} can serve as a powerful probe of the chirality and spin polarization of the edge state, and thus of the topological order \cite{wenbook} in a QH liquid.

\section{Model}

\begin{figure}
\centering
\includegraphics[width=3.2in]{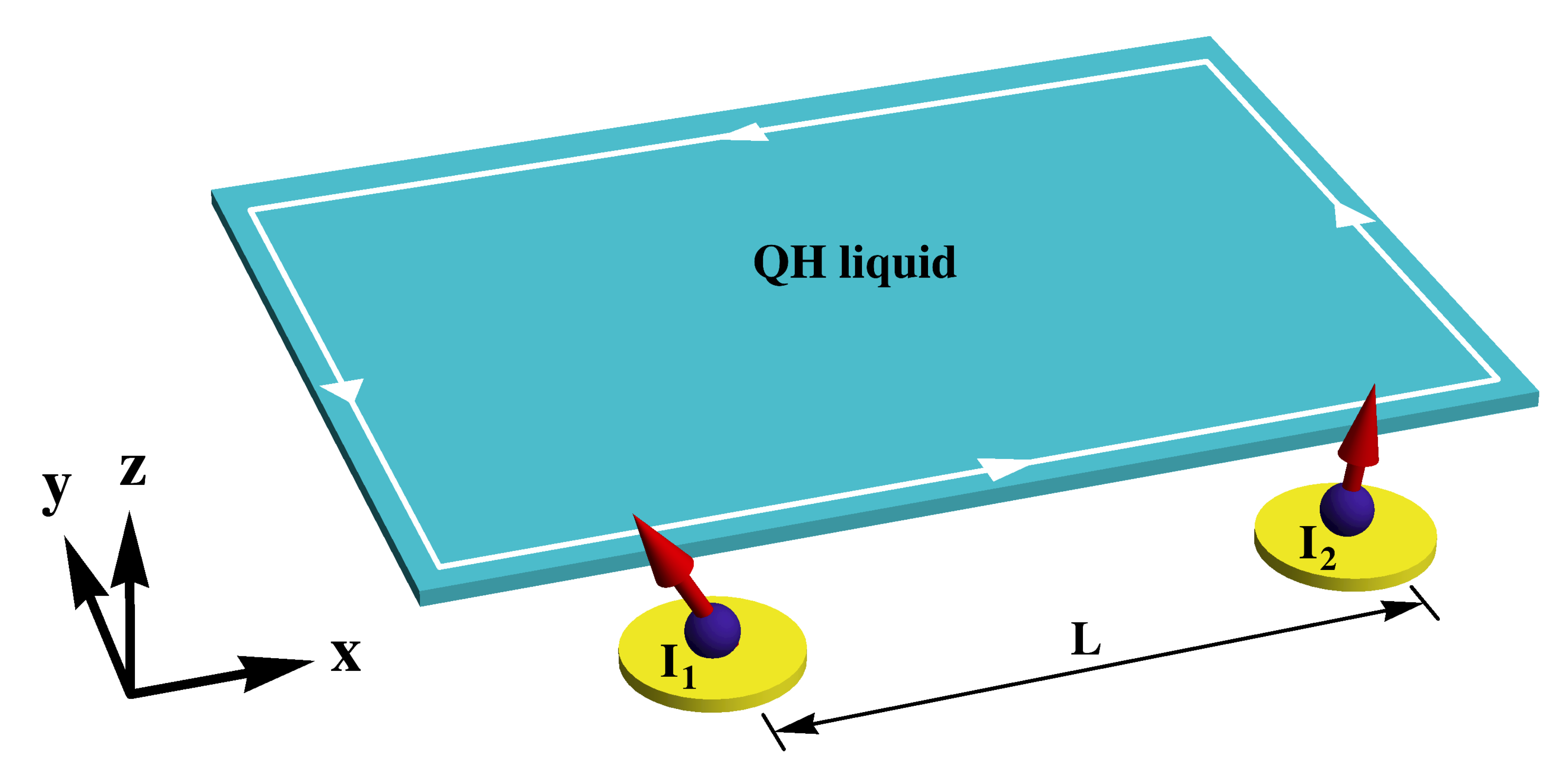}
\caption{The basic setup consisting of two QDs (yellow disks) tunnel coupled to the edge (white lines and arrows) of a QH liquid (blue sheet) confined in the $x$-$y$ plane. In general, the QH edge may support multiple edge modes, propagating in the same or opposite direction(s), which we do not depict explicitly. The QDs are separated by a distance $L$ along the QH edge. Each QD contains a single electron (blue spheres), whose spin (red arrows) serves as a qubit. The coupling strength between the QDs and the QH edge is controlled by gates (not shown). We assume no direct interaction between the localized electron spins $\mathbf{I}_1$ and $\mathbf{I}_2$ in the QDs.}
\end{figure}

We now discuss the physics of RKKY interaction mediated by QH edge states. The basic setup is shown in Fig.~1. Two QDs are placed adjacent to a QH liquid, separated by a distance $L$ and labelled by the site index $i=1,2$. Conduction electrons in the QH edge state can tunnel into and out of the QDs \cite{kiyama15} and thus can interact with the localized spins in them. This establishes coupling between the QH edge and the QDs. For simplicity, we treat the QDs as two spatial points. The Hamiltonian describing such a system has the form 
\begin{equation}
\label{m1}
H=H_{\textrm{edge}}+\sum_{i=1,2} \Gamma_i \mathbf{S}_i \cdot \mathbf{I}_i,
\end{equation}
where $H_{\textrm{edge}}$ is the Hamiltonian of conduction electrons in the edge state, $\mathbf{I}_i=(I_i^x,I_i^y,I_i^z)$ denotes the localized spin in the $i$th QD, and $\mathbf{S}_i=(S_i^x,S_i^y,S_i^z)$ denotes the spin of conduction electrons coupled to $\mathbf{I}_i$, with coupling strength $\Gamma_i$. Experimentally, $\Gamma_i$ can be tuned by gating. We define $\mathbf{S}_i$ to be the spin density in the edge state multiplied by the confinement length of the QDs. For the setup, we assume $L$ is large so that there is no direct interaction between the spins in the QDs. 

In the weak tunnel coupling regime such that $\Gamma_i\ll E_F$, where $E_F$ is the Fermi energy of conduction electrons, the dynamics of the spins in the QDs effectively decouples from that of the conduction electrons. In such a case, one can derive an effective Hamiltonian for the spins in QDs, valid in the adiabatic regime, by performing a Schrieffer-Wolff transformation \cite{sw,bravyi11} of Eq.~(\ref{m1}) followed by tracing out the degrees of freedom of conduction electrons (see Appendix A for the derivation of effective Hamiltonian and a discussion of adiabaticity), 
\begin{equation}
\label{m2}
H_{\textrm{eff}}=\sum_{ij,\alpha\beta}J_{ij}^{\alpha\beta} I^{\alpha}_i   I^{\beta}_j-\sum_{i} \mathbf{B}_i \cdot \mathbf{I}_i,
\end{equation}
where the spin-component indices $\alpha,\beta=x,y,z$. 
The first term is the RKKY interaction, with $J_{ij}^{\alpha\beta}=\Gamma_i\Gamma_j\chi_{ij}^{\alpha\beta}/2$. Here $\chi_{ij}^{\alpha\beta}$ is the static spin susceptibility of conduction electrons, $\chi_{ij}^{\alpha\beta}=-i \int^{\infty}_0dt \ e^{-\eta t} \langle [S^{\alpha}_i(t),S^{\beta}_j(0)]\rangle$, where $\eta=0^+$ and $\langle\cdots\rangle$ denotes the average determined by $H_{\textrm{edge}}$. Physically, conduction electrons in the vicinity of a QD develop a spin-density oscillation due to their interaction with the spin in the QD. This spin-density response, determined by $\chi_{ij}^{\alpha\beta}$, can be perceived by the spins in other QDs coupled to the QH edge. In this way the RKKY interaction is established. For spin-unpolarized QH states, we assume $\langle S_i^x\rangle=\langle S_i^y\rangle=\langle S_i^z\rangle=0$, such that $\chi_{ij}^{\alpha\beta}=\delta^{\alpha\beta}\chi_{ij}^{\alpha\alpha}$. On the other hand, the in-plane spin operators $S_i^x$, $S_i^y$ are less relevant (in the renormalization group sense) than the out-of-plane ones $S_i^z$ in a QH state with full spin polarization, as we discuss below.  In this case we set $S_i^x=S_i^y=0$ and hence $\chi_{ij}^{\alpha\beta}=\delta^{\alpha z}\delta^{\beta z}\chi_{ij}^{zz}$. Thus, in general we have $J_{ij}^{\alpha\beta}=\delta^{\alpha\beta}J_{ij}^{\alpha\alpha} $. The RKKY interaction leads to an effective exchange coupling $\mathcal{J}^{\alpha}=J^{\alpha\alpha}_{12}+J^{\alpha\alpha}_{21}$, as a function of the interdot distance $L$, between the localized spins $ I^{\alpha}_1$ and $ I^{\alpha}_2$. The effective onsite Zeeman fields $\mathbf{B}_i$ are a direct consequence of time-reversal (TR) symmetry breaking in QH systems. We find $\mathbf{B}_i=(\Gamma_i^2/2)\int_{0}^{\infty}dt \ e^{-\eta t}\langle \mathbf{S}_i(t)\times \mathbf{S}_i(0)\rangle$. In spin-polarized QH states, $\mathbf{B}_i=0$ (for more details and estimates we refer to Appendix \ref{App_A_Effective Hamiltonian}).

\section{RKKY interaction in various QH states}

The RKKY interaction in Eq.~(\ref{m2}) is by nature long-ranged and can be used as an approach to entangle spin qubits over long distances. Thus, it is important to understand how the interaction looks like in various QH systems. To this end, it is convenient to adopt a continuum description of the QH edge states that  is well approximated by the chiral Luttinger liquid (LL) model at low energy \cite{wenbook}. In general, the edge of a QH liquid may support (electron-) density-fluctuation modes as well as Majorana fermions (zero-modes), with the action
\begin{align}
\label{m3}
S_{\textrm{edge}}=&\int dxdt \ \Big[
 \sum_{IJ}\frac{1}{4\pi}(K_{IJ}\partial_t\phi_I\partial_x\phi_J -V_{IJ}\partial_x\phi_I\partial_x\phi_J) \nonumber \\
&+  \sum_K i\lambda_K (\partial_{t}-v_{K}\partial_{x})\lambda_K \Big],
\end{align}
written in the bosonization language \cite{wenbook} (throughout the article we set $\hbar=1$). The bosonic fields $\phi_I$ describe the density modes, and $\lambda_K$ denote the Majorana fermions. The symmetric matrix $K_{IJ}$ encodes the topological properties of the QH state, while the positive-definite symmetric matrix $V_{IJ}$ specifies the velocities and interactions of $\phi_I$. The parameter $v_K$ is the velocity of $\lambda_K$: $v_K>0$ ($v_K<0$) if the $\lambda_K$ is left-moving (right-moving).

Upon passing to the continuum limit, we replace the spin operators $S^{\alpha}_i(t)/l$ with spin density operators $S^{\alpha}(x_i,t)$, where $l$ is the confinement length of the QDs and $x_i$ is the position of the $i$th QD. The nonvanishing components of the spin susceptibility are given by $\chi^{\alpha\alpha}_{ij}=-il^2 \int^{\infty}_0dt\ e^{-\eta t} \langle [S^{\alpha}(x_i,t),S^{\alpha}(x_j,0)]\rangle$. Assuming translation invariance along the QH edge, which is justified for clean samples, we may further write $\chi^{\alpha\alpha}_{ij}=\chi^{\alpha\alpha}(x_i-x_j)$, \cite{gbook} where
\begin{equation}
\label{m4}
\chi^{\alpha\alpha}(x)=2 l^2\int^{\infty}_0dt\ e^{-\eta t} \textrm{Im}\langle \mathcal{T} S^{\alpha}(x,t)S^{\alpha}(0,0)\rangle ,
\end{equation}
with $\mathcal{T}$ the time-ordering operator. The correlators are evaluated in the zero-temperature limit.
We define
\begin{equation}
\label{m5}
S^{\alpha}(x,t)=\frac{1}{2}\sum_{\sigma\sigma'}\psi_{\sigma}^{\dagger} (x,t)\sigma^{\alpha}_{\sigma\sigma'}\psi_{\sigma'} (x,t),
\end{equation}
where $\psi_{\sigma}=\sum_{\mu}\psi^{\mu}_{\sigma}$ is the sum of the most-relevant electron operators $\psi^{\mu}_{\sigma}$ with spin $\sigma=\uparrow,\downarrow$ on the QH edge. 

The number of $\psi^{\mu}_{\uparrow}$ operators is not necessarily equal to that of $\psi^{\mu}_{\downarrow}$ operators since TR symmetry is broken. For instance, the most-relevant electron operators have the same spin in a spin-polarized QH state, so that $S^x=S^y=0$. This is in contrast to the situation in 1D systems where TR symmetry is present \cite{braunecker09,lee15}. Using bosonization, we express $S^{\alpha}$ in terms of the fields $\phi_I$ and $\lambda_K$, and compute the spin susceptibility.

We sketch the calculation of the spin susceptibility for a generic QH edge state (for particular examples, see Appendix B). First of all, we assume separation of charged and neutral degrees of freedom in the QH edge state. This phenomenon, as has been demonstrated experimentally in a number of QH systems \cite{bid10,bocquillon13}, results from strong Coulomb interaction among the elementary density modes $\phi_I$ and resembles ``charge-spin separation'' in a generic TR-invariant 1D system \cite{gbook}. As a result, the physical modes that propagate on the QH edge are the charged and neutral collective modes as well as Majorana fermions. The physical parameters relevant to experiment are the velocities and interactions of these propagating modes, whose magnitudes are set by different energy scales in the QH system. For instance, the charged-mode velocity, determined by the dominant Coulomb energy scale, is much greater than the velocity of neutral mode and other parameters~\cite{bocquillon13}. We make use of this fact in our calculation. For a moment, we consider the case of two density modes in the edge theory, see Eq.~(\ref{m3}). To compute the correlators in Eq.~(\ref{m4}), we define a new set of fields which diagonalize the action of the density modes $\phi_I$. 
The action takes the form
\begin{align}
S_{\textrm{density}}=&\int dxdt\ \frac{1}{4\pi}\Big[\partial_t\phi_{+}\partial_x\phi_{+} +\varepsilon  \partial_t\phi_{-}\partial_x\phi_{-} \nonumber \\
&  -v_{+}\partial_x\phi_{+}\partial_x\phi_{+} -v_{-}\partial_x\phi_{-}\partial_x\phi_{-}], 
\end{align}
in the basis of new fields $\phi_+$ and $\phi_-$. Here $\varepsilon=1$ ($\varepsilon=-1$)  if the edge states are chiral (non-chiral) and $v_+,v_->0$.  New velocities $v_+$ and $v_-$ are well approximated by the velocities of the physical charged mode and neutral mode, respectively, so that $v_+ \gg v_-$. Upon expressing the spin density operators in terms of the free fields $\phi_+$, $\phi_-$, and $\lambda_K$, it is straightforward to compute the correlators, 
\begin{align}
\langle \mathcal{T} S^{\alpha}(x,t)&S^{\alpha}(0,0)\rangle  \propto   \cos(\Delta kx) \Big[\frac{1}{\delta+i(t+x/v_+)}\Big]^{g_+^{\alpha}} \nonumber\\
& \times \Big[\frac{1}{\delta+i(t+\varepsilon x/v_-)}\Big]^{g_-^{\alpha}} , 
\end{align}
where $\delta>0$ is an infinitesimal and $\Delta k$ is the gauge-invariant momentum difference between the edge modes. The case $\Delta k=0$ corresponds to the scattering of an edge mode with itself. Here we have omitted the terms that are less relevant, and assumed $|v_K|=v_-$ as both of the velocities are determined by less dominant energy scales in the system. The exponents $g_+^{\alpha}$, $g_-^{\alpha}$ are functions of the matrices $K_{IJ}$ and $V_{IJ}$ and as we show  $0<g_+^{\alpha}\ll 1$ and $g_-^{\alpha}> 1$ (see Appendix B for the expressions in different QH states). Evaluating the time integral in Eq.~(\ref{m4}), we obtain $\chi^{\alpha\alpha}(x)$, which in general may contain multiple terms for different momentum differences. We keep only the most-relevant terms.

The various QH states can be divided into three types: (i) Those with a chiral edge state containing a single density mode, such as the Laughlin states at filling factors $\nu=1/m$, where $m$ is an odd integer. (ii) Those with a chiral edge state containing multiple interacting density modes, such as the QH state at $\nu=2$. (iii) Those with a non-chiral edge state, such as the particle-hole dual states \cite{girvin84} of Laughlin states. 

For QH states of type (i), we find $\chi^{\alpha\alpha}(x)=0$, taking into account the most-relevant spin operators in the edge state. Thus, to the lowest order the RKKY interaction cannot be established. Physically, the vanishing spin susceptibility reflects the homogeneous electronic structure in an independent QH edge mode, a property originating from the incompressibility of the QH liquid which prevents the formation of electronic spin texture. In reality, however, a small nonzero spin susceptibility may still be measured, due to higher-order processes involving virtual transitions to edge states in higher Landau levels. 

In QH edge states of type (ii) and type (iii), the spin susceptibility is nonzero to the lowest order. In these cases, the inter-edge interactions introduce inhomogeneous degrees of freedom (``noise'') to the stream of conduction electrons, allowing for the development of spin-density oscillations. We find
\begin{equation}
\label{m6}
\chi^{\alpha\alpha}(x)= \frac{\cos(\Delta kx)}{|x|^{g^{\alpha}}} \Theta(- x) C^{\alpha} (\boldsymbol{g}^{\alpha}, \boldsymbol{v}),
\end{equation}
for left-moving type (ii) edge states, where $g^{\alpha}=g_+^{\alpha}+g_-^{\alpha}-1$,  $\Theta(x)$ is the Heaviside step function, and $C^{\alpha} (\boldsymbol{g}^{\alpha}, \boldsymbol{v})$ are functions of $\boldsymbol{g}^{\alpha}=(g_+^{\alpha},g_-^{\alpha})$ and $\boldsymbol{v}=(v_+,v_-)$, whose explicit definitions are given in Appendix B. 
If the edge state is right-moving, one replaces $\Theta(- x)$ with $\Theta(x)$, and sends $\boldsymbol{v}\rightarrow -\boldsymbol{v}$ in $C^{\alpha} (\boldsymbol{g}^{\alpha}, \boldsymbol{v})$. These findings suggest that the spin susceptibility in type (ii) edge states is ``rectified'', {\it i.e.}, directed in the down-stream direction of the propagation of conduction electrons, see Fig.~2(a), where left- and right-moving directions are defined with respect to the lower edge of the QH liquid (the same in Fig.~2(b)). This result is not surprising and can be understood also intuitively. In a left-moving edge state, conduction electrons move in the $-x$ direction, leading to the factor $\Theta (-x)$ in the expression of $\chi^{\alpha\alpha}(x)$. Formally, such an interesting form of the spin susceptibility is a manifestation of the causality principle in 1D chiral systems, where information is transported one-way and novel physical rules can emerge, e.g., see Ref.~\cite{fdt} for fluctuation-dissipation relations in chiral QH systems. 

Lastly, we find
\begin{align}
\label{m7}
\chi^{\alpha\alpha}(x)=\frac{\cos(\Delta kx)}{|x|^{g^{\alpha}}}  \{\Theta(x) C^{\alpha} _>(\boldsymbol{g}^{\alpha}, \boldsymbol{v})  +\Theta(-x) C^{\alpha} _<(\boldsymbol{g}^{\alpha}, \boldsymbol{v})  \},
\end{align}
for type (iii) edge states, where $C^{\alpha}_> (\boldsymbol{g}^{\alpha}, \boldsymbol{v})$ and $C^{\alpha}_< (\boldsymbol{g}^{\alpha}, \boldsymbol{v})$ are functions of $\boldsymbol{g}^{\alpha}$ and $\boldsymbol{v}$, defined in Appendix B. The spin susceptibility in this case is ``both-way'', as shown in Fig.~2(b), with different magnitudes in the $+x$ and $-x$ directions, i.e., $C^{\alpha}_> (\boldsymbol{g}^{\alpha}, \boldsymbol{v})\neq C^{\alpha}_< (\boldsymbol{g}^{\alpha}, \boldsymbol{v})$. This again reflects the inequivalence between left-moving and right-moving edge modes.  Imagining now the chirality of all edge modes are reverted, e.g., by TR operation, the profile of the spin susceptibility should also be reverted. Indeed, we find that $C^{\alpha}_> (\boldsymbol{g}^{\alpha}, \boldsymbol{v})$ are related to $C^{\alpha}_< (\boldsymbol{g}^{\alpha}, \boldsymbol{v})$ by the exchange of arguments  $v_+\leftrightarrow v_-$ and $g_+^{\alpha}\leftrightarrow g_-^{\alpha}$, which technically carries out the chirality-reverting procedure (see Appendix B). In the above discussion, we have assumed that spin excitations do not extend into the $L_0-L$ part of the QH edge, where $L_0$ is the total edge length. In practice, this is realized by grounding the  $L_0-L$ part or by choosing the sample such that $L_0\gg L$.

\begin{figure}
\centering
\subfloat[]{\includegraphics[width=1.6in]{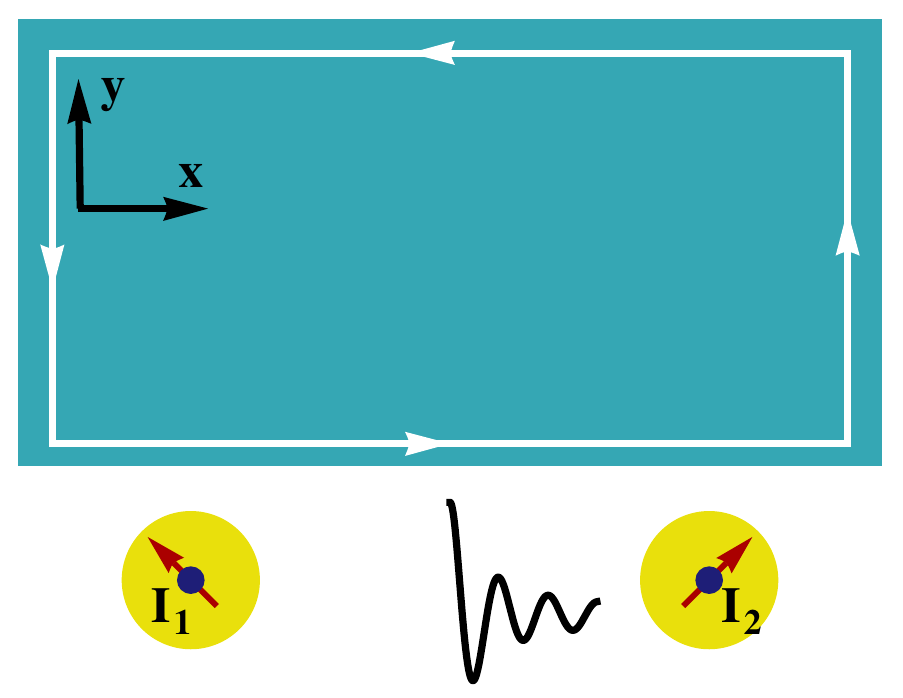}}
\quad
\subfloat[]{\includegraphics[width=1.6in]{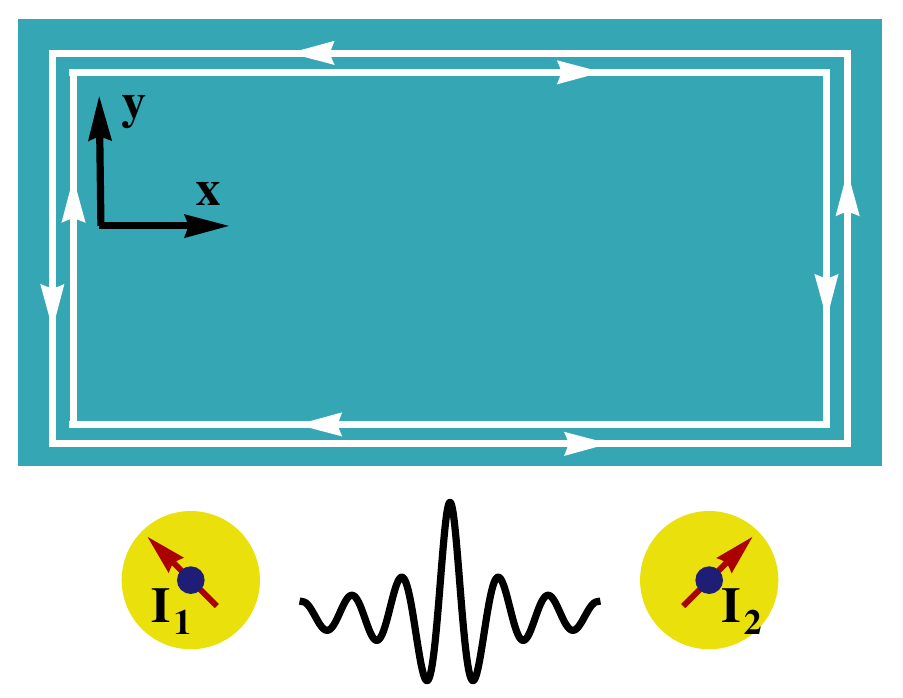}}
\caption{Spin susceptibility in QH edge states of (a) type (ii) and (b) type (iii). For the type (ii) case, the spin susceptibility is directed in the propagation direction of the edge modes. For the type (iii) case, the spin susceptibility is nonzero in both directions along the QH edge.}
\end{figure}

The exponents $g^{\alpha}$, where $\alpha=x,y,z$, determine how the RKKY interaction scales with distance. In Table~\ref{tableI}, we list them in different QH states. In general, $g^{\alpha}$ depend on both the chirality and spin polarization of the QH edge state. For chiral edge states, i.e., those of type (ii), these exponents are integral invariants depending on the topological order of the bulk liquid, whereas for non-chiral edge states they are nonuniversal and depend on the parameters in the Hamiltonian. In the latter case, we write  $g^{\alpha}=g^{\alpha}_0+\delta g^{\alpha}$, where $g^{\alpha}_0$ is the integer part of $g^{\alpha}$. As shown in Appendix B, $\delta g^{\alpha}/g^{\alpha}_0\ll 1$ for all the non-chiral edge states in the table, assuming ``charge-neutral separation'' on the edge. Moreover, we find that the in-plane components of the RKKY interaction vanish in a spin-polarized QH state, leading to an Ising-type exchange coupling of spin qubits. On the other hand, the RKKY interaction has zero out-of-plane component and equal in-plane components in a spin-unpolarized QH state, which is XY-type. This suggests that a transformation of the anisotropy type of the RKKY interaction may be observed in the QH liquid at $\nu=2/3$, which was found to be spin-unpolarized at low fields and spin-polarized at high fields \cite{eisenstein90}. 

The QH state at $\nu=5/2$ is also of special interest. We consider both Abelian and non-Abelian topological orders proposed to describe this state. The former include the Halperin $331$ state \cite{halperin83} and 113 state \cite{yang14}, and the latter include the Moore-Read (Pfaffian) state \cite{moore91}, the anti-Pfaffian state \cite{levin07,lee07} and the $SU(2)_2$ state \cite{wen91}. The $331$ state and $113$ state can be both spin-polarized and -unpolarized, just like the $\nu=2/3$ QH state. 
The Pfaffian state, like the Laughlin states, supports a single density mode on the edge and thus has vanishing RKKY interaction. The particle-hole dual state of the Pfaffian state, the anti-Pfaffian state, has a non-chiral edge state and a non-integer scaling exponent. For the $SU(2)_2$ state, we assume that the Majorana fermion and the neutral collective mode propagate at different velocities, as they should in reality, which is necessary to obtain a nonvanishing scaling exponent. Such careful treatment is not essential for other $\nu=5/2$ states. We have assumed that the RKKY interaction is mediated solely by the fractional edge modes in the second Landau level, while the integer edge modes in the lowest Landau level do not play a role. Experimentally, this can be fulfilled, using the fact that edge modes in different Landau levels are spatially separated \cite{csg}. For instance, the QDs in Fig.~1 can be moved out of the plane of the QH liquid and formed in a second two- or quasi-one-dimensional electron gas in the vertical direction~\cite{Eisenstein1990,Auslaender2000,Wegscheider1993,scheller14}, such that they are in tunnel contact with the fractional edge modes but far away from the integer edge modes. 
The coupling between the integer edge and the QDs and the interaction between the integer edge and the fractional edge can be neglected to a good approximation.

 \begin {table*}[t]
\begin{center}
    \begin{tabular}{ | c | c | c | c | c || c | c | c |c|c|c| c|}
    \hline
   ~QH state ~~&~~$\overline{1/m}$~~&~~$2$~~&~~$2/3$~~&~~$\overline{2/3}$~~&~~331~~&~~$\overline{331}$~~&~~113~~&~~$\overline{113}$~~&~~$\overline{\textrm{Pf}}$~~&~~$\overline{\textrm{APf}}$~~&~~$\overline{SU(2)_2}$~~\\ \hline
    $g^x$ & $-$ & $\overrightarrow{1}$ & $ 1$&$-$ & $\overrightarrow{3}$ & $-$&$ 3$ &$-$ & $-$&$-$ & $-$\\ \hline
    $g^y$ & $-$ & $\overrightarrow{1}$ & $ 1$&$-$ & $\overrightarrow{3}$ & $-$& $ 3$&$-$ & $-$&$-$ & $-$\\ \hline
    $g^z$ & $-$ & $-$ & $-$& $ 1$ & $-$ & $\overrightarrow{3}$ &$-$ & $ 3$& $-$&$ 1$ & $\overrightarrow{1}$\\ \hline
  RKKY type   &  & XY & XY& Ising & XY & Ising &XY & Ising& &Ising & Ising\\ \hline
    \end{tabular}
\end{center}
\caption {Scaling exponents $g^{\alpha}$ and anisotropy type of the RKKY interaction in various QH states. An overline is used to indicate a spin-polarized state, e.g., $\overline{2/3}$ denotes the spin-polarized QH state at $\nu=2/3$. We consider several topological orders at $\nu=5/2$, including both Abelian ones (the 331 state and the 113 state, denoted as 331/$\overline{331}$ and 113/$\overline{113}$, respectively) and non-Abelian ones (the Pfaffian state, the anti-Pfaffian state, and the $SU(2)_2$ state, denoted as $\overline{\textrm{Pf}}$, $\overline{\textrm{APf}}$, and $\overline{SU(2)_2}$, respectively). The 331 state and the 113 state both have spin-unpolarized and spin-polarized versions. For chiral edge states, the exponents are integers and we add arrows to indicate that the spin susceptibility is non-zero only in the down-stream direction.  For non-chiral edge states, the exponents are non-integers and we enter the integer parts $g^{\alpha}_0$ of the exponents. We put ``$-$'' in the entry if the corresponding component of the spin susceptibility (and thus that of the RKKY interaction) vanishes. The RKKY interaction is XY-type in spin-unpolarized states and Ising-type in spin-polarized states.
}
\label{tableI}
\end {table*}

\section{Discussion}

Let us estimate the coupling between the two spin qubits in Fig.~1, given by $\mathcal{J}^{\alpha}=\Gamma_1\Gamma_2\{\chi^{\alpha\alpha}(L)+\chi^{\alpha\alpha}(-L)\}/2$. In Appendix B, we obtain the dimensional part $[\chi^{\alpha\alpha}(x)]$ of the spin susceptibility,
\begin{align}
\label{m8}
[\chi^{\alpha\alpha}(x)]\simeq l^2a^{g^{\alpha}-1}|x|^{-g^{\alpha}}/v_-,
\end{align}
for both type (ii) and type (iii) edge states, where $a$ is the lattice constant of the underlying material hosting the QH system. 
For example, let us consider the QH state at $\nu=2$, realized in GaAs heterostructures. We have $a= 0.565$~nm for GaAs, $g^x=g^y=1$, and $v_-\simeq 10^{4}$~m/s \cite{bocquillon13}. Using $\Gamma_1=\Gamma_2=\Gamma\simeq 0.1$~meV and $l=30$~nm (see Appendix C for the estimates), we find $\mathcal{J}^{x}=\mathcal{J}^{y}\simeq1~\mu$eV for $L=1~\mu$m. This is about one order of magnitude smaller than the direct exchange strength $\mathcal{J}_{\textrm{direct}}\simeq 10-100~\mu$eV in typical GaAs double QDs \cite{loss98} and is experimentally measurable. The RKKY interaction established by QH edge states thus provides a way to realize entangled quantum gates over mesoscopic distances. The implementation of two-qubit gates using Hamiltonians of the form of Eq.~(\ref{m2}) is well known: see, e.g., Ref.~\cite{loss98} (footnote 13) for Ising-type coupling and Ref.~\cite{imamoglu99} for XY-type coupling. The $\mu$eV exchange strength converts to gate-operation times of the order of nanoseconds, which is well below the coherence times \cite{qcreview} of spin qubits. 

It is interesting to compare the RKKY interaction in QH edge states with that in semiconductor quantum wires. Assuming spin-rotation symmetry, the dimensional part $[\chi_{\textrm{w}}(x)]$ of the spin susceptibility in quantum wires can be found in Ref.~\cite{braunecker09}. The ratio 
\begin{align}
\label{m9}
r^{\alpha}(x)=\frac{[\chi^{\alpha\alpha}(x)]}{[\chi_{\textrm{w}}(x)]}=\frac{v_F}{v_-}\Big(\frac{a}{|x|} \Big)^{g^{\alpha}-g_{\textrm{w}}} 
\end{align}
characterizes the relative strength of the RKKY interaction in the two sorts of systems, where $v_F$ is the Fermi velocity in the quantum wire and $g_{\textrm{w}}$ depends on the interaction of electrons. In non-interacting case, $g_{\textrm{w}}=1$. Consider the $\nu=2$ QH edge state and GaAs quantum wire. We find $r^x(L)=r^y(L)\simeq 1.5$ for $L=1~\mu$m, using $g_{\textrm{w}} = 0.75$ and $v_F\simeq 10^{5}~$m/s \cite{braunecker09}. In principle, quantum wires can also be used to mediated RKKY interaction between spin qubits. However, using QH edge states offers more advantages. From technical aspect, the edge states and the spin qubits can be realized in the same material, for instance, in a 2DEG in GaAs heterostructures, which is more experimentally accessible than a setup with quantum wires. More importantly, the topologically protected QH edge states are more immune to disorder effects and perturbations in the system than quantum wires. This guarantees a better quality of the long-distance quantum gates. 

Our discussions so far have focused on the RKKY interaction between spin qubits. Interestingly, the treatments can also be applied to obtain the RKKY interaction between nuclear spins embedded in the 1D QH edge state (see also Ref.~\cite{meng14}). 
To this end, let $\Gamma_i\rightarrow A/N$ and $\mathbf{I}_i\rightarrow\tilde{\mathbf{I}}_i$ in Eq.~(\ref{m1}) and the following equations, where $A$ is the hyperfine coupling constant, $N$ is the number of nuclear spins in a cross section (labelled by $i$) of the QH edge, and $\tilde{\mathbf{I}}_i$ is the total nuclear spin operator in a given cross section. Given a non-chiral edge state with both spin-up and spin-down electrons, e.g., the spin-unpolarized state at $\nu=2/3$, the nuclear spins may form a helical magnetic order \cite{braunecker09} at low temperatures, induced by the RKKY interaction. The nuclear magnetic order acts back on the electronic system by gapping out conducting edge modes. Experimentally, such an order is evidenced by the reduction of the conductance at low temperatures \cite{scheller14}.

By measuring the spatial dependence of RKKY interaction \cite{stano,rkkyexp}, one can obtain information about the chirality and spin polarization of the QH edge state, which in turn are related to the topological order of the bulk QH liquid \cite{wenbook}. In particular, this technique may be used to detect the nature of the QH liquid at $\nu=5/2$: One can distinguish between a chiral edge state and a non-chiral edge state by confirming whether the spin susceptibility is unidirectional along the edge. One can rule out either a spin-polarized state or a spin-unpolarized state by comparing the in-plane and out-of-plane components of the RKKY interaction, by measuring the spin states in the QDs. For this one can make use of experimental techniques based on spin-to-charge conversion \cite{loss98} developed for read-out of spin qubits in QDs \cite{Petta,Hanson,Marcus}. 
The numerical values of the scaling exponents also help to identify the true $\nu=5/2$ state. The advantages of measuring the spin susceptibility are obvious, compared with other approaches detecting topological orders based on edge-bulk correspondence \cite{wenbook}, such as measuring the temperature and voltage dependence of quasiparticle tunneling \cite{wenqp}. First, it is easier to vary the sampling point in space than in temperature or voltage, e.g., one may use the setup in Fig.~1 with an array of QDs. Second, information encoded in spin degrees of freedom is more robust than that encoded in charge current, against unfavorable modification due to long-range Coulomb interaction in the device \cite{yang13}. Compared with electronic Fabry-P\'{e}rot \cite{stern06, bonderson06} and Mach-Zehnder \cite{feldman06,wang10,yang15} interferometries, our setup probes the non-Abelian topological orders at $\nu=5/2$ with a much simpler device geometry and more straightforward data. 

The scenario becomes more complicated if one replaces the QDs with quantum anti-dots \cite{geller97}. 
In that case, tunneling of quasiparticles, rather than electrons, defines the coupling between the QH edge and the anti-dots. It is still possible to define an RKKY interaction mediated by quasiparticles in the edge state, whose spatial dependence can be used to distinguish different Abelian QH states. For non-Abelian states, however, there are ambiguities in the scaling behavior of the RKKY interaction, arising from the multiple fusion channels of non-Abelian quasiparticles. 

To conclude, we have introduced a novel approach to achieving long-distance entanglement of spin qubits confined in QDs, based on the RKKY interaction mediated by QH edge states. The approach allows for the implementation of quantum gates with long coupling ranges and fast operation times, which would greatly facilitate the development of large-scale quantum computers. From fundamental point of view, the ability to probe the chirality and the spin polarization of a QH edge state via measuring the spatial form of the RKKY interaction opens up a new venue for studying electronic and spin physics in QH systems.

\acknowledgments
We acknowledge support from the Swiss NSF and NCCR QSIT.

\appendix

\section{Effective Hamiltonian}
\label{App_A_Effective Hamiltonian}

Our starting point is the Hamiltonian in  Eq.~(\ref{m1}). For weak tunnel coupling between the QH edge and the QDs, we can treat $H_{\Gamma}=\sum_{i} \Gamma_i \mathbf{S}_i \cdot \mathbf{I}_i
$ as a perturbation and make a Schrieffer-Wolff transformation~\cite{sw,bravyi11} to remove terms linear in $\Gamma_i$ from the Hamiltonian. The transformed Hamiltonian reads
\begin{equation}
\label{a1}
\bar{H}=e^SHe^{-S}=H_{\textrm{edge}}-\frac{1}{2}[[H_{\textrm{edge}},S],S]+\cdots,
\end{equation}
where $S$ satisfies $[H_{\textrm{edge}},S]=H_{\Gamma}$. Written in terms of the Liouvillian superoperator $\mathcal{L}$, $S=\mathcal{L}^{-1}H_{\Gamma}$. The leading-order terms in $\Gamma_i$ in $\bar{H}$ are given by 
\begin{equation}
\label{a2}
\bar{H}_{\Gamma}=-\frac{1}{2}[[H_{\textrm{edge}},S],S]=\frac{1}{2}[\mathcal{L}^{-1}H_{\Gamma},H_{\Gamma}].
\end{equation}
Using $\mathcal{L}^{-1}=-i\int_0^{\infty}dt\ e^{-\eta t}e^{i\mathcal{L}t}$, where $\eta=0^+$, we find
\begin{align}
\label{a3}
\bar{H}_{\Gamma}=&-\frac{i}{2}\int_0^{\infty}dt\ e^{-\eta t}[H_{\Gamma}(t),H_{\Gamma}] \nonumber \\
=&-\frac{i}{2}\sum_{ij}\Gamma_i\Gamma_j\int_0^{\infty}dt\ e^{-\eta t} [\mathbf{S}_i(t) \cdot \mathbf{I}_i,\mathbf{S}_j(0) \cdot \mathbf{I}_j] \nonumber \\
=&-\frac{1}{2}\int_0^{\infty}dt\ e^{-\eta t} \{\sum_{ij}i\Gamma_i\Gamma_j  I_i^{\alpha}I_j^{\beta}[S_i^{\alpha}(t),S_j^{\beta}(0)] \nonumber\\
&+\sum_{i}\Gamma_i^2\epsilon^{\alpha\beta\gamma}I_i^{\alpha}S_i^{\beta}(t)S_i^{\gamma}(0)\},
\end{align}
where we have defined $\hat{O}(t)=e^{iH_{\textrm{edge}}t}\hat{O}e^{-iH_{\textrm{edge}}t}$ for an operator $\hat{O}$ and used $[I_i^{\alpha},I_j^{\beta}]=i\delta_{ij}\epsilon^{\alpha\beta\gamma}I_i^{\gamma}$, with $\epsilon^{\alpha\beta\gamma}$ the Levi-Civita symbol. Summation over repeated spin-component indices (Greek letters) is implied throughout this appendix. 

Next, we take the expectation $\langle \cdots\rangle$ over the electronic degrees of freedom in the QH edge state. This gives an effective Hamiltonian describing the dynamics of localized spins in the adiabatic limit,
\begin{equation}
\label{a4}
H_{\textrm{eff}}=\langle\bar{H}_{\Gamma}\rangle=\sum_{ij}\frac{\Gamma_i\Gamma_j}{2} \chi_{ij}^{\alpha\beta}I_i^{\alpha}I_j^{\beta}-\sum_{i} B_i^{\alpha}  I_i^{\alpha}, 
\end{equation}
where we have identified the spin susceptibility of conduction electrons,
\begin{equation}
\label{a5}
\chi_{ij}^{\alpha\beta}=-i \int^{\infty}_0dt\ e^{-\eta t} \langle [S^{\alpha}_i(t),S^{\beta}_j(0)]\rangle,
\end{equation}
and defined effective onsite Zeeman fields for the QDs,
\begin{equation}
\label{a6}
B_i^{\alpha}=\frac{\Gamma_i^2}{2}\int_{0}^{\infty}dt\ e^{-\eta t}\epsilon^{\alpha\beta\gamma}\langle S_i^{\beta}(t)S_i^{\gamma}(0)\rangle.
\end{equation}
This is the Hamiltonian in Eq.~(\ref{m2}).

In deriving the above effective Hamiltonian, we have neglected the external magnetic field $B^{\rm ext}$ that leads to the formation of the QH liquid. To justify this, let us estimate $B_i^{\alpha}$ and $B^{\rm ext}$ (in unit of energy). For spin-unpolarized QH states, let us assume that the two terms in Eq.~(\ref{a3}) have the same order of magnitude after taking the expectation and performing the integration, for typical time scales related to the dynamics of conduction electrons. This gives $[B_i^{\alpha}]\sim \Gamma_i^2[\chi_{ii}^{\alpha\alpha}]$, where $[\cdots]$ denotes the dimensional part. Passing to the continuum limit, let $[\chi_{ii}^{\alpha\alpha}]\rightarrow [\chi^{\alpha\alpha}(a)]$, where $[\chi^{\alpha\alpha}(x)]$ is given by Eq.~(\ref{m8}) and $a$ is the natural short-distance cut-off, taken as the lattice constant of the host material. An estimation similar to that for the effective exchange coupling $\mathcal{J}^{\alpha}$ finds $B_i^{\alpha}\simeq 1$~meV. Meanwhile, $B^{\rm ext}\simeq 0.1$~meV for typical field strengths of several Tesla in QH liquids. Thus, $B_i^{\alpha}$ is large compared with $B^{\rm ext}$ in spin-unpolarized states.

For spin-polarized QH states, applying the assumption $S_i^x=S_i^y=0$ yields $ B_i^{\alpha}=0$. In this case, we consider fluctuations in the next order, associated with the next-most-relevant spin operators $\delta S_i^x,\delta S_i^y$ in the edge theory. We have $\langle\delta S_i^x\rangle=\langle \delta S_i^y\rangle=0$. The fluctuations give rise to effective onsite Zeeman fields 
\begin{equation}
\label{a7}
\delta B_i^{\alpha}=\frac{\Gamma_i^2}{2}\int_{0}^{\infty}dt\ e^{-\eta t}\epsilon^{\alpha\beta\gamma}\langle \delta S_i^{\beta}(t)\delta S_i^{\gamma}(0)\rangle,
\end{equation}
which are fully out-of-plane, $\delta B_i^{\alpha}=\delta^{\alpha z} \delta B_i^z$. In other words, the spin-polarized edge states tend to polarize the spin qubits. Simple dimensional analysis shows that the order of magnitude of $\delta S_i^x,\delta S_i^y $ differs from those of the (nonvanishing) most-relevant spin operators by a factor of $a/\bar{v} \tau$, where $\bar{v}$ is the mean edge velocity and $\tau\sim 1/E_F$ is a typical time scale for the dynamics of conduction electrons. Accordingly, the factor $(a/\bar{v} \tau)^2 $ enters the relative strength of the effective onsite fields ($ B_i^{\alpha}$) in spin-unpolarized states to that ($\delta B_i^{\alpha}$) in spin-polarized states (where $ B_i^{\alpha}=0$). Our estimation shows that $a/\bar{v} \tau <0.1$, so that $\delta B_i^{\alpha}\ll B^{\rm ext}\ll B_i^{\alpha}$. In the main text we have neglected $\delta B_i^{\alpha}$ for simplicity.

In principle, the Zeeman terms $H_Z=- \sum_i B^{\rm ext}I_i^z$ (assuming $\mathbf{B}^{\rm ext}=B^{\rm ext}\hat{z}$) should be included in the unperturbed Hamiltonian in the Schrieffer-Wolff procedure, i.e., $H_{\textrm{edge}}\rightarrow H_{\textrm{edge}}+H_Z$ in Eqs.~(\ref{a1}) and (\ref{a2}) and the definition of time evolution. As a consequence, the first localized-spin operator $I^{\alpha}_i$ appearing in the two terms in Eq.~(\ref{a3}) acquires time dependence, in addition to the time dependence in the first conduction-spin operator $S^{\alpha}_i$. The dynamics of $I^{\alpha}_i$, set by the Zeeman energy $B^{\rm ext}$, however decouples from that of $S^{\alpha}_i$, set by the Fermi energy $E_F$, since $E_F \gg B^{\rm ext}$ according to the estimation above. Thus, to a good approximation we may neglect the time dependence in $I^{\alpha}_i$. We do this for spin-unpolarized states. For spin-polarized states, Eq.~(\ref{a3}) is exact: Only the terms with $\alpha=\beta=z$ survive in the equation and we have  $I^{\alpha}_i(t)=I^{\alpha}_i(0)$ since $H_Z$ commutes with $I^z_i$. 

We note moreover that $H_Z$ also appears in Eq.~(\ref{a4}) for both spin-unpolarized and spin-polarized QH states. In the main text, we have neglected this term for simplicity. However, $H_Z$ must be taken into account for the purpose of implementing two-qubit quantum gates. 

The effective Hamiltonian in Eq.~(\ref{a4}) describes the system in Fig.~1 in equilibrium. Given a change in the spin state of one of the qubits, the entire electronic system readjusts to achieve new equilibrium. The change of the  qubit must be adiabatic in order for the other qubit to sense the change and respond. This means that the switching time $t_{\rm sw}$ of the first qubit satisfies $t_{\rm sw}\gg L/\bar{v}$. On the other hand, if the qubit state is changed very fast (non-adiabatically), there will be no effect on the second qubit within time $L/\bar{v}$. In that case, the process is dynamic and is described by the spin susceptibility at finite frequencies. For $L=1~\mu$m, $L/\bar{v} \simeq 10$~ps, which is much shorter than the ideal gate-operation time $t_{\rm sw}\simeq 1$~ns. Thus, the requirement for adiabaticity does not place much restriction on the operation of spin qubits.

\section{Spin Susceptibility}

In this appendix, we calculate the spin susceptibility for the QH states listed in Table~I. The formula is given by Eq.~(\ref{m4}). First, we compute the correlators  in the zero-temperature limit
\begin{equation}
\mathcal{G}^{\alpha}(x,t)=\langle \mathcal{T} S^{\alpha}(x,t)S^{\alpha}(0,0)\rangle,
\end{equation}
where $\alpha=x,y,z$. We focus on the scaling behaviors of these correlators and neglect the proportionality constants. Next, we evaluate the time integral
\begin{equation}
\label{b0}
\chi^{\alpha\alpha}(x)=2 l^2\int^{\infty}_0dt\ e^{-\eta t} \textrm{Im}\mathcal{G}^{\alpha}(x,t) ,
\end{equation}
where $\eta=0^+$. Restoring the proportionality constants, we obtain the full expression of the spin susceptibility.

\subsection{Correlators}

\subsubsection{Laughlin states at $\nu=1/m$}
The Lagrangian density that describes the edge state of the $\nu=1/m$ ($m$ is an odd integer) Laughlin state is 
\begin{equation}
\label{b1}
L=\frac{m}{4\pi}[\partial_t \phi \partial_x \phi - v(\partial_x \phi )^2],
\end{equation}
where $v$ is the velocity of the edge mode described by bosonic field $\phi$.
We assume the edge state is left-moving. Electrons in the edge state are described by the vertex operator $\psi= \frac{1}{\sqrt{2\pi a}}e^{-i  k_Fx}e^{-i m\phi}$, where $a$ is the short-distance cut-off and $k_F$ is the Fermi momentum. Here and throughout this appendix we omit the Klein factors in the electron operators, which will drop out when evaluating the average. Since the edge state is spin-polarized, all the electrons have the same spin $\sigma$. Let us assume $\sigma=\uparrow$. Using Eq.~(\ref{m5}) and neglecting transitions to higher Landau levels, we find $S^x=S^y=0$, and
\begin{align}
S^z=\frac{1}{2}\psi^{\dagger}\psi \propto \partial_x \phi.
\end{align}
The correlator of $\phi$ can be read from Eq.~(\ref{b1}), $\langle\mathcal{T}\phi(x,t)\phi(0,0)\rangle=-\nu\ln(x+vt-i\delta)+{\rm const.}$, where $\delta$ is defined as a positive infinitesimal throughout the appendix. 
This gives
\begin{equation}
\label{addb3}
\mathcal{G}^{z}(x,t)\propto \frac{\nu}{(x+vt-i \delta)^2},
\end{equation}
whereas $\mathcal{G}^{x}=\mathcal{G}^{y}=0$. Substituting Eq.~(\ref{addb3}) in Eq.~(\ref{b0}) we obtain the spin susceptibility in Laughlin states.

\subsubsection{The QH state at $\nu=2$}
The $\nu=2$ QH state has two bosonic edge modes $\phi_\uparrow,\phi_\downarrow$, propagating in the same direction, where $\phi_\uparrow$ has spin up and $\phi_\downarrow$ has spin down. The Lagrangian density is 
\begin{equation}
L=\frac{1}{4\pi}\{\sum_{i=\uparrow,\downarrow}[\partial_t \phi_i \partial_x \phi_i- v_i(\partial_x \phi_i )^2]-2u\partial_x \phi_{\uparrow}\partial_x \phi_{\downarrow}\},
\end{equation}
where $v_i$ is the velocity of $\phi_i$ and $u>0$ is the repulsive Coulomb interaction between $\phi_\uparrow$ and $\phi_\downarrow$. We assume the edge modes are left-moving.
The most-relevant electron operators are
$\psi_i=\frac{1}{\sqrt{2\pi a}}e^{-i k_{F,i}x}e^{-i \phi_i}$, where $k_{F,i}$ is the Fermi momentum of $\phi_i$.
The spin density operators are
\begin{align}
S^x&=\frac{1}{2}(\psi^{\dagger}_{\uparrow}\psi_{\downarrow}+\textrm{H.c.}) \propto e^{i\Delta k x}e^{i(\phi_{\uparrow}-\phi_{\downarrow})}+\textrm{H.c.}\nonumber \\
S^y&=\frac{1}{2}(-i\psi^{\dagger}_{\uparrow}\psi_{\downarrow}+\textrm{H.c.})  \propto-ie^{i\Delta k x}e^{i(\phi_{\uparrow}-\phi_{\downarrow})}+\textrm{H.c.}\nonumber \\
S^z&=\frac{1}{2}(\psi^{\dagger}_{\uparrow}\psi_{\uparrow}-\psi^{\dagger}_{\downarrow}\psi_{\downarrow})\propto \partial_x( \phi_{\uparrow}-\phi_{\downarrow}), 
\end{align}
where $\Delta k=k_{F,\uparrow}-k_{F,\downarrow}$ is the gauge-invariant momentum difference, proportional to the magnetic flux penetrating between the two edge modes. 

To compute the correlators, we define eigenmodes
\begin{align}
\phi_+&=  \cos{\varphi}~\phi_{\uparrow} +\sin{\varphi}~\phi_{\downarrow} \nonumber\\ 
\phi_-&=  -\sin{\varphi}~\phi_{\uparrow} +\cos{\varphi}~\phi_{\downarrow}  ,
\end{align}
where $\tan{2\varphi}=\frac{2u}{v_{\uparrow}-v_{\downarrow}}$, which diagonalize the edge theory,
\begin{equation}
L=\frac{1}{4\pi}\sum_{i=+,-}[\partial_t \phi_i \partial_x \phi_i- v_i(\partial_x \phi_i )^2],
\end{equation}
where $v_{\pm}=\frac{1}{2}(v_{\uparrow}+v_{\downarrow}\pm \sqrt{(v_{\uparrow}-v_{\downarrow})^2+4u^2})$. According to the experiment \cite{bocquillon13}, $v_+\gg v_-$ as a result of the strong Coulomb interaction $u$. Expressing the spin density operators in eigenmodes, it is straightforward to obtain
\begin{align}
\mathcal{G}^{x}(x,t)\propto &\cos(\Delta kx) \Big[\frac{1}{x+v_+t-i  \delta}\Big]^{c_+^2}\Big[\frac{1}{x+v_-t-i  \delta}\Big]^{c_-^2} \nonumber \\
\mathcal{G}^{y}(x,t)\propto &\cos(\Delta kx) \Big[\frac{1}{x+v_+t-i  \delta}\Big]^{c_+^2}\Big[\frac{1}{x+v_-t-i  \delta}\Big]^{c_-^2} \nonumber \\
\mathcal{G}^{z}(x,t)\propto &\frac{c_+^2}{(x+v_+t-i \delta)^2}+\frac{c_-^2}{(x+v_-t-i \delta)^2},
\end{align}
where the functions $c_{\pm}(\varphi)=\cos{\varphi} \mp \sin{\varphi}$. 
Notice that $c_+^2(\varphi)+c_-^2(\varphi)=2$, i.e., the scaling exponents of the correlators are integral invariant, independent of the angle $\varphi$ which depends on the inter-edge interaction. This is a well-known property of chiral QH edge states \cite{wenbook}.

\subsubsection{The QH state at $\nu=2/3$}
The $\nu=2/3$ QH state can be spin-unpolarized at low fields and spin-polarized at high fields~\cite{eisenstein90}.

We first consider the spin-unpolarized state. It has two bosonic edge modes $\phi_{\uparrow}$ and $\phi_{\downarrow}$, where $\phi_{\uparrow}$ has spin up and $\phi_{\downarrow}$ has spin down. The Lagrangian density is
\begin{equation}
\label{b2}
L= \frac{1}{4\pi}  \sum_{i,j=\uparrow,\downarrow} [K_{ij}\partial_{t}\phi_i\partial_{x}\phi_j-V_{ij}\partial_{x}\phi_i\partial_{x}\phi_j],
\end{equation}
where
\begin{equation}
K =
\left( \begin{array}{cc}
 1 & 2 \\
 2 & 1 \end{array} \right)
\mspace{9.0mu} \textrm{and} \mspace{9.0mu}
V =
\left( \begin{array}{cc}
 v_{\uparrow} & u \\
 u & v_{\downarrow} \end{array} \right),
\end{equation}
with $v_i$ the velocity of $\phi_i$ and $u$ the inter-edge interaction. The eigenvalues of the $K$-matrix have opposite signs, so the edge state is non-chiral.

Experiment \cite{bid10} revealed that the $\nu=2/3$ edge state consists of a charged mode and a neutral mode, moving in opposite directions. To connect the parameters in the edge theory described by Eq. (\ref{b2}) with experiment, we change to the physical basis of charged mode $\phi_{\rho}=\phi_{\uparrow}+\phi_{\downarrow}$ and neutral mode $\phi_{n}=\phi_{\uparrow}-\phi_{\downarrow}$,
\begin{align}
\label{b3}
L =  \frac{1}{4\pi}  & [ \frac{3}{2}\partial_{t}\phi_{\rho}\partial_{x}\phi_{\rho}-\frac{1}{2}\partial_{t}\phi_{n}\partial_{x}\phi_{n} - \frac{3}{2} v_{\rho} (\partial_{x}\phi_{\rho})^{2} \nonumber \\
 &-\frac{1}{2}v_{n} (\partial_{x}\phi_{n})^{2}-2v_{\rho n}\partial_{x}\phi_{\rho}\partial_{x}\phi_{n} ] ,
\end{align}
where $v_{\rho}=\frac{1}{3}(\frac{v_{\uparrow}}{2}+\frac{v_{\downarrow}}{2}+u)$, $v_n=\frac{v_{\uparrow}}{2}+\frac{v_{\downarrow}}{2}-u$, and $v_{\rho n}=\frac{1}{4}(v_{\uparrow}-v_{\downarrow})$. In general, $v_{\uparrow}\neq v_{\downarrow}$ due to finite Zeeman splitting. The charged-mode velocity $v_{\rho}$, determined by the large Coulomb energy scale, is expected to be much greater in order of magnitude than the neutral-mode velocity $v_n$ and the interaction $v_{\rho n}$. We therefore assume $v_{\rho}\gg v_n \sim v_{\rho n}$. In particular, we assume that the scaling dimensions of quasiparticle operators in the real case do not deviate much from those in the case $v_{\rho n}=0$. With this assumption, we can determine the most-relevant electron operators in the edge theory, which are $\psi_{\uparrow}\propto e^{-i (2k_{F,\uparrow}+k_{F,\downarrow})x}e^{-i (2\phi_{\uparrow}+\phi_{\downarrow})}$, with spin up, and $\psi_{\downarrow}\propto e^{-i (k_{F,\uparrow}+2k_{F,\downarrow})x}e^{-i (\phi_{\uparrow}+2\phi_{\downarrow})}$, with spin down, where $k_{F,\uparrow}$ and $k_{F,\downarrow}$ are momentum-like constants related to the spatial locations of the edge modes $\phi_{\uparrow}$ and $\phi_{\downarrow}$.
The spin density operators are obtained by computing the operator product expansions (OPEs) of the electron operators and keeping the most-singular terms. We find   
\begin{align}
S^x&=\frac{1}{2}(\psi^{\dagger}_{\uparrow}\psi_{\downarrow}+\textrm{H.c.}) \propto e^{i\Delta k x}e^{i\phi_n}+\textrm{H.c.}  \nonumber \\
S^y&=\frac{1}{2}(-i\psi^{\dagger}_{\uparrow}\psi_{\downarrow}+\textrm{H.c.}) \propto -ie^{i\Delta k x}e^{i\phi_n}+\textrm{H.c.} \nonumber \\
S^z&=\frac{1}{2}(\psi^{\dagger}_{\uparrow}\psi_{\uparrow}-\psi^{\dagger}_{\downarrow}\psi_{\downarrow}) \propto \partial_x\phi_n,
\end{align}
where $\Delta k=k_{F,\uparrow}-k_{F,\downarrow}$.

In terms of eigenmodes 
\begin{align}
\label{b4}
\phi_+ &= \sqrt{\frac{3}{2}} \cosh{\theta}~\phi_{\rho} +\sqrt{\frac{1}{2}}\sinh{\theta}~\phi_n \nonumber\\ 
\phi_- & = \sqrt{\frac{3}{2}} \sinh{\theta}~\phi_{\rho} +\sqrt{\frac{1}{2}}\cosh{\theta}~\phi_n ,
\end{align}
where $\tanh{2\theta}=\frac{4}{\sqrt{3}}\frac{v_{\rho n}}{v_{\rho}+v_{n}}$, the edge theory is diagonalized,
\begin{equation}
L=\frac{1}{4\pi}[\partial_t \phi_+ \partial_x \phi_+ -\partial_t \phi_- \partial_x \phi_- - \sum_{i=+,-} v_i(\partial_x \phi_i )^2],
\end{equation}
where $v_+=\frac{1}{\cosh{2\theta}}(\cosh^2{\theta}v_{\rho}-\sinh^2{\theta}v_{n})$ and $v_-=\frac{1}{\cosh{2\theta}}(\cosh^2{\theta}v_{n}-\sinh^2{\theta}v_{\rho})$. Since $v_{\rho}\gg v_n \sim v_{\rho n}$, we have $ \theta \ll 1$ and thus $v_+\simeq v_{\rho}$, $v_-\simeq v_n$ and $v_+\gg v_-$. 
The correlators are evaluated to be 
\begin{align}
\mathcal{G}^{x}(x,t) &\propto \cos(\Delta kx) \Big[\frac{1}{x+v_+t-i  \delta}\Big]^{\tilde{c}_+^2}\Big[\frac{1}{x-v_-t+i  \delta}\Big]^{\tilde{c}_-^2} \nonumber \\
\mathcal{G}^{y}(x,t)&\propto \cos(\Delta kx) \Big[\frac{1}{x+v_+t-i  \delta}\Big]^{\tilde{c}_+^2}\Big[\frac{1}{x-v_-t+i \delta}\Big]^{\tilde{c}_-^2} \nonumber \\
\mathcal{G}^{z}(x,t)&\propto \frac{\tilde{c}_+^2}{(x+v_+t-i \delta)^2}+\frac{\tilde{c}_-^2}{(x-v_-t+i \delta)^2},
\end{align}
where the functions $\tilde{c}_+(\theta)=\sqrt{2}\sinh{\theta}$ and $\tilde{c}_-(\theta)=\sqrt{2}\cosh{\theta}$. Notice that $\tilde{c}_+^2(\theta)+\tilde{c}_-^2(\theta)=2(1+2\sinh^2{\theta})$, i.e., the scaling exponents are non-universal and depend on the parameters in the Hamiltonian, through $\theta$. This reflects the non-chiral nature of the edge state.

Next, we discuss the spin-polarized state at $\nu=2/3$. It has two bosonic edge modes $\phi_1$ and $\phi_2$, having the same spin polarization (assuming they are spin-up). The Lagrangian density has the same form of Eq.~(\ref{b2}), with 
\begin{equation}
K =
\left( \begin{array}{cc}
 1 & 0 \\
 0 & -3 \end{array} \right)
\mspace{9.0mu} \textrm{and} \mspace{9.0mu}
V =
\left( \begin{array}{cc}
 v_1 & u' \\
 u' & 3v_2 \end{array} \right).
\end{equation}
This is also a non-chiral state. The charged mode and the neutral mode in the edge theory are identified as  $\phi_{\rho}=\phi_{1}+\phi_{2}$ and $\phi_{n}=\phi_{1}+3\phi_{2}$, respectively, in terms of which the Lagrangian density recovers the expression in Eq.~(\ref{b3}), with $v_{\rho}=\frac{3}{2}v_1+\frac{1}{2}v_2-u'$, $v_n=\frac{1}{2}v_1+\frac{3}{2}v_2-u'$ and $v_{\rho n}=-\frac{3}{4}v_1-\frac{3}{4}v_2+u'$. Again we assume $v_{\rho}\gg v_n \sim v_{\rho n}$. The most-relevant electron operators are $\psi_1\propto  e^{-i (2k_{F,1}+3k_{F,2})x}e^{-i (2\phi_1+3\phi_2)}$ and $\psi_2\propto  e^{-i k_{F,1}x}e^{-i \phi_1}$, both with spin up, where $k_{F,1}$ and $k_{F,2}$ are constants. Using Eq.~(\ref{m5}) and OPE, we find $S^x=S^y=0$, and $S^z=S^z_f+S^z_b$, where
\begin{align}
S^z_f&=\frac{1}{2}(\psi^{\dagger}_1\psi_1+\psi^{\dagger}_2\psi_2)\propto \partial_x\phi_{\rho}\nonumber\\
S^z_b&=\frac{1}{2}(\psi^{\dagger}_1\psi_2+\textrm{H.c.})\propto e^{i\Delta k x}e^{i\phi_n}+\textrm{H.c.},
\end{align}
where $\Delta k=k_{F,1}+3k_{F,2}$ is interpreted as the Fermi-momentum difference between the elementary edge mode $\phi_1$ and $\phi_2$. The rest of the analysis resembles that for the spin-unpolarized state. We diagonalize the edge theory using the free fields $\phi_+,\phi_-$ defined in Eq.~(\ref{b4}) and evaluate the correlators. We find $\mathcal{G}^{x}=\mathcal{G}^{y}=0$, and $\mathcal{G}^{z}=\mathcal{G}^{z}_f+\mathcal{G}^{z}_b$, where 
\begin{align}
\mathcal{G}^{z}_f(x,t) &\propto \frac{\tilde{c}_-^2}{(x+v_+t-i \delta)^2}+\frac{\tilde{c}_+^2}{(x-v_-t+i \delta)^2} \nonumber \\
\mathcal{G}^{z}_b(x,t) &\propto \cos(\Delta kx) \Big[\frac{1}{x+v_+t-i  \delta}\Big]^{\tilde{c}_+^2}\Big[\frac{1}{x-v_-t+i  \delta}\Big]^{\tilde{c}_-^2}.
\end{align}

\subsubsection{The 331 state at $\nu=5/2$}

We now turn to the QH state at $\nu=5/2$. This QH state is usually modeled by combining a $\nu=2$ integer QH state in the lowest Landau level, which is treated as an inert background assuming no Landau level mixing, and a $\nu=1/2$ fractional QH state in the second Landau level, which is assumed to capture the full topological order of the QH liquid. We study the RKKY interaction mediated solely by the fractional edge state. In the following, we consider several topological orders proposed for the fractional edge state, including Halperin's 331 and 113 states \cite{halperin83,yang14}, the Pfaffian state \cite{moore91}, the anti-Pfaffian state \cite{levin07,lee07} and the $SU(2)_2$ state \cite{wen91}. Motivated by the experiment \cite{bid10}, we will always assume separation of charged and neutral degrees of freedom in the edge state. Moreover, we assume that the charged-mode velocity is much greater than other physical parameters, by a similar argument to that for the QH state at $\nu=2/3$. 

We start from Halperin's 331 state, which has a spin-unpolarized version and a spin-polarized version. The Lagrangian density for the edge of the spin-unpolarized 331 state has the same form of Eq.~(\ref{b2}), with
\begin{equation}
K =
\left( \begin{array}{cc}
 3 & 1\\
 1 & 3 \end{array} \right)
\mspace{9.0mu} \textrm{and} \mspace{9.0mu}
V =
\left( \begin{array}{cc}
 v_{\uparrow} & u \\
 u & v_{\downarrow} \end{array} \right),
\end{equation}
where $v_{\uparrow}$ and $v_{\downarrow}$ are the velocities of edge modes $\phi_{\uparrow}$ and $\phi_{\downarrow}$, respectively, and $u$ the inter-edge interaction. Here $\phi_{\uparrow}$ is a spin-up mode and $\phi_{\downarrow}$ is a spin-down mode. The 331 state is chiral.
The physical charged mode and neutral mode are defined as $\phi_{\rho}=\phi_{\uparrow}+\phi_{\downarrow}$ and $\phi_{n}=\phi_{\uparrow}-\phi_{\downarrow}$, respectively, in terms of which the Lagrangian density is
\begin{align}
\label{b5}
L =  \frac{1}{4\pi} & [ 2\partial_{t}\phi_{\rho}\partial_{x}\phi_{\rho}+\partial_{t}\phi_{n}\partial_{x}\phi_{n} - 2 v_{\rho} (\partial_{x}\phi_{\rho})^{2} \nonumber\\
 &-v_{n} (\partial_{x}\phi_{n})^{2}-2v_{\rho n}\partial_{x}\phi_{\rho}\partial_{x}\phi_{n} ] ,
\end{align}
where $v_{\rho}=\frac{v_{\uparrow}}{8}+\frac{v_{\downarrow}}{8}+\frac{u}{4}$, $v_n=\frac{v_{\uparrow}}{4}+\frac{v_{\downarrow}}{4}-\frac{u}{2}$ and $v_{\rho n}=\frac{1}{4}(v_{\uparrow}-v_{\downarrow})$. Assuming $v_{\rho}\gg v_n \sim v_{\rho n}$, the most-relevant electron operators are $\psi_{\uparrow}\propto e^{-i (3k_{F,\uparrow}+k_{F,\downarrow})x}e^{-i (3\phi_{\uparrow}+\phi_{\downarrow})}$, with spin up, and $\psi_{\downarrow}\propto e^{-i (k_{F,\uparrow}+3k_{F,\downarrow})x}e^{-i (\phi_{\uparrow}+3\phi_{\downarrow})}$, with spin down, where $k_{F,\uparrow}$ and $k_{F,\downarrow}$ are constants. The spin density operators are
\begin{align}
\label{b6}
S^x&=\frac{1}{2}(\psi^{\dagger}_{\uparrow}\psi_{\downarrow}+\textrm{H.c.}) \propto e^{i\Delta k x}e^{i2\phi_n}+\textrm{H.c.}\nonumber \\
S^y&=\frac{1}{2}(-i\psi^{\dagger}_{\uparrow}\psi_{\downarrow}+\textrm{H.c.}) \propto -ie^{i\Delta k x}e^{i2\phi_n}+\textrm{H.c.}\nonumber \\
S^z&=\frac{1}{2}(\psi^{\dagger}_{\uparrow}\psi_{\uparrow}-\psi^{\dagger}_{\downarrow}\psi_{\downarrow})\propto \partial_x\phi_n ,
\end{align}
where $\Delta k=2k_{F,\uparrow}-2k_{F,\downarrow}$.
To evaluate the correlators of $S^{\alpha}$, we define eigenmodes
\begin{align}
\label{b7}
\phi_+&= \sqrt{2} \cos{\theta}~\phi_{\rho} +\sin{\theta}~\phi_{n} \nonumber\\ 
\phi_-&= -\sqrt{2} \sin{\theta}~\phi_{\rho} +\cos{\theta}~\phi_{n} ,
\end{align}
where $\tan{2\theta}=\frac{\sqrt{2}v_{\rho n}}{v_{\rho}-v_{n}}$. We have $ \theta \ll 1$. We find 
\begin{align}
\mathcal{G}^{x}(x,t)&\propto \cos(\Delta kx) \Big[\frac{1}{x+v_+t-i  \delta}\Big]^{\bar{c}_+^2}\Big[\frac{1}{x+v_-t-i  \delta}\Big]^{\bar{c}_-^2} \nonumber \\
\mathcal{G}^{y}(x,t)&\propto \cos(\Delta kx) \Big[\frac{1}{x+v_+t-i  \delta}\Big]^{\bar{c}_+^2}\Big[\frac{1}{x+v_-t-i  \delta}\Big]^{\bar{c}_-^2}\nonumber \\
\mathcal{G}^{z}(x,t)&\propto \frac{\bar{c}_+^2}{(x+v_+t-i \delta)^2}+\frac{\bar{c}_-^2}{(x+v_-t-i \delta)^2},
\end{align}
where $\bar{c}_+(\theta)=2\sin{\theta}$ and $\bar{c}_-(\theta)=2\cos{\theta}$. The parameters $v_+$ and $v_-$ are the velocities of $\phi_+$ and $\phi_-$, respectively. We have $v_+\simeq v_{\rho}$, $v_-\simeq v_n$ and $v_+\gg v_-$.

The spin-polarized 331 state has two bosonic edge modes $\phi_1$ and $\phi_2$, having the same spin polarization (assuming they are spin-up). The Lagrangian density has the same form of Eq.~(\ref{b2}), with
\begin{equation}
K =
\left( \begin{array}{cc}
 3 & -2 \\
 -2 & 4 \end{array} \right)
\mspace{9.0mu} \textrm{and} \mspace{9.0mu}
V =
\left( \begin{array}{cc}
 v_1 & u' \\
 u' & v_2 \end{array} \right).
\end{equation}
 The physical charged mode and neutral mode are identified as $\phi_{\rho}=\phi_1$ and  $\phi_{n}=-\phi_1+2\phi_2$, respectively, in terms of which the Lagrangian density recovers the form in Eq.~(\ref{b5}), with $v_{\rho}=\frac{1}{2}v_1+\frac{1}{8}v_2+\frac{1}{2}u'$, $v_n=\frac{1}{4}v_2$ and $v_{\rho n}=\frac{1}{4}v_2+\frac{1}{2}u'$. Assuming $v_{\rho}\gg v_n \sim v_{\rho n}$, the most-relevant electron operators are $\psi_1\propto e^{-i (k_{F,1}+2k_{F,2})x} e^{-i (\phi_1+2\phi_2)}$ and $\psi_2\propto e^{-i (3k_{F,1}-2k_{F,2})x}e^{-i (3\phi_1-2\phi_2)}$, both with spin up, where $k_{F,1}$ and $k_{F,2}$ are constants. The spin density operators are $S^x=S^y=0$, and $S^z=S^z_f+S^z_b$, where
\begin{align}
S^z_f&=\frac{1}{2}(\psi^{\dagger}_1\psi_1+\psi^{\dagger}_2\psi_2)\propto \partial_x\phi_{\rho}\nonumber\\
S^z_b&=\frac{1}{2}(\psi^{\dagger}_1\psi_2+\textrm{H.c.})\propto e^{i\Delta k x}e^{i2\phi_n}+\textrm{H.c.},
\end{align}
with $\Delta k=-2k_{F,1}+4k_{F,2}$. Using the definition of eigenmodes in Eq.~(\ref{b7}), we find $\mathcal{G}^{x}=\mathcal{G}^{y}=0$, and $\mathcal{G}^{z}=\mathcal{G}^{z}_f+\mathcal{G}^{z}_b$, where 
\begin{align}
\mathcal{G}^{z}_f(x,t) &\propto \frac{\bar{c}_-^2}{(x+v_+t-i \delta)^2}+\frac{\bar{c}_+^2}{(x+v_-t-i \delta)^2} \nonumber \\
\mathcal{G}^{z}_b(x,t) &\propto\cos(\Delta kx) \Big[\frac{1}{x+v_+t-i  \delta}\Big]^{\bar{c}_+^2}\Big[\frac{1}{x+v_-t-i  \delta}\Big]^{\bar{c}_-^2}.
\end{align}

\subsubsection{The 113 state at $\nu=5/2$}
The 113 state also has a spin-unpolarized version and a spin-polarized version. 
 
The edge theory of the spin-unpolarized 113 state is of the form of Eq.~(\ref{b2}), with
\begin{equation}
K =
\left( \begin{array}{cc}
 1 & 3\\
 3 & 1 \end{array} \right)
\mspace{9.0mu} \textrm{and} \mspace{9.0mu}
V =
\left( \begin{array}{cc}
 v_{\uparrow} & u \\
 u & v_{\downarrow} \end{array} \right),
\end{equation}
where $v_{\uparrow}$ and $v_{\downarrow}$ are the velocities of edge modes $\phi_{\uparrow}$ and $\phi_{\downarrow}$, respectively, and $u$ the inter-edge interaction. Here $\phi_{\uparrow}$ is a spin-up mode and $\phi_{\downarrow}$ is a spin-down mode. The 113 state is non-chiral. Switching to the physical basis of charged mode $\phi_{\rho}=\phi_{\uparrow}+\phi_{\downarrow}$ and neutral mode $\phi_{n}=\phi_{\uparrow}-\phi_{\downarrow}$, the Lagrangian density becomes
\begin{align}
\label{b8}
L =  \frac{1}{4\pi} & [ 2\partial_{t}\phi_{\rho}\partial_{x}\phi_{\rho}-\partial_{t}\phi_{n}\partial_{x}\phi_{n} - 2 v_{\rho} (\partial_{x}\phi_{\rho})^{2} \nonumber\\
& -v_{n} (\partial_{x}\phi_{n})^{2}-2v_{\rho n}\partial_{x}\phi_{\rho}\partial_{x}\phi_{n} ] ,
\end{align}
where $v_{\rho}=\frac{v_{\uparrow}}{8}+\frac{v_{\downarrow}}{8}+\frac{u}{4}$, $v_n=\frac{v_{\uparrow}}{4}+\frac{v_{\downarrow}}{4}-\frac{u}{2}$ and $v_{\rho n}=\frac{1}{4}(v_{\uparrow}-v_{\downarrow})$. Assuming $v_{\rho}\gg v_n \sim v_{\rho n}$, the most-relevant electron operators are $\psi_{\uparrow}\propto e^{-i (3k_{F,\uparrow}+k_{F,\downarrow})x}e^{-i (3\phi_{\uparrow}+\phi_{\downarrow})}$, with spin up, and $\psi_{\downarrow}\propto e^{-i (k_{F,\uparrow}+3k_{F,\downarrow})x}e^{-i (\phi_{\uparrow}+3\phi_{\downarrow})}$, with spin down, where $k_{F,\uparrow}$ and $k_{F,\downarrow}$ are constants. The spin density operators are
\begin{align}
\label{b6}
S^x&=\frac{1}{2}(\psi^{\dagger}_{\uparrow}\psi_{\downarrow}+\textrm{H.c.}) \propto e^{i\Delta k x}e^{i2\phi_n}+\textrm{H.c.}\nonumber \\
S^y&=\frac{1}{2}(-i\psi^{\dagger}_{\uparrow}\psi_{\downarrow}+\textrm{H.c.}) \propto -ie^{i\Delta k x}e^{i2\phi_n}+\textrm{H.c.}\nonumber \\
S^z&=\frac{1}{2}(\psi^{\dagger}_{\uparrow}\psi_{\uparrow}-\psi^{\dagger}_{\downarrow}\psi_{\downarrow})\propto \partial_x\phi_n ,
\end{align}
where $\Delta k=2k_{F,\uparrow}-2k_{F,\downarrow}$. The eigenmodes are defined as 
\begin{align}
\label{b9}
\phi_+&= \sqrt{2} \cosh{\theta}~\phi_{\rho} +\sinh{\theta}~\phi_{n} \nonumber\\ 
\phi_-&= \sqrt{2} \sinh{\theta}~\phi_{\rho} +\cosh{\theta}~\phi_{n} ,
\end{align}
where $\tanh{2\theta}=\frac{\sqrt{2}v_{\rho n}}{v_{\rho}+v_{n}}$. We have $ \theta \ll 1$. We find
\begin{align}
\mathcal{G}^x(x,t)& \propto \cos(\Delta kx) \Big[\frac{1}{x+v_+t-i  \delta}\Big]^{2\tilde{c}_+^2}\Big[\frac{1}{x-v_-t+i  \delta}\Big]^{2\tilde{c}_-^2} \nonumber \\
\mathcal{G}^y(x,t)& \propto \cos(\Delta kx) \Big[\frac{1}{x+v_+t-i  \delta}\Big]^{2\tilde{c}_+^2}\Big[\frac{1}{x-v_-t+i  \delta}\Big]^{2\tilde{c}_-^2} \nonumber \\
\mathcal{G}^z(x,t)& \propto \frac{\tilde{c}_+^2}{(x+v_+t-i \delta)^2}+\frac{\tilde{c}_-^2}{(x-v_-t+i \delta)^2},
\end{align}
where $\tilde{c}_+(\theta)=\sqrt{2}\sinh{\theta}$ and $\tilde{c}_-(\theta)=\sqrt{2}\cosh{\theta}$. The parameters $v_+$ and $v_-$ are the velocities of $\phi_+$ and $\phi_-$, respectively. We have $v_+\simeq v_{\rho}$, $v_-\simeq v_n$ and $v_+\gg v_-$.

The spin-polarized 113 state has two bosonic edge modes $\phi_1$ and $\phi_2$, having the same spin polarization (assume they are spin-up). The Lagrangian density has the same form of Eq.~(\ref{b2}), with
\begin{equation}
K =
\left( \begin{array}{cc}
 1 & 2 \\
 2 & -4 \end{array} \right)
\mspace{9.0mu} \textrm{and} \mspace{9.0mu}
V =
\left( \begin{array}{cc}
 v_1 & u' \\
 u' & v_2 \end{array} \right).
\end{equation}
The charged mode and the neutral mode are $\phi_{\rho}=\phi_1$ and  $\phi_{n}=-\phi_1+2\phi_2$, respectively, in terms of which the Lagrangian density recovers the form in Eq.~(\ref{b8}), with $v_{\rho}=\frac{1}{2}v_1+\frac{1}{8}v_2+\frac{1}{2}u'$, $v_n=\frac{1}{4}v_2$ and $v_{\rho n}=\frac{1}{4}v_2+\frac{1}{2}u'$. Assuming $v_{\rho}\gg v_n \sim v_{\rho n}$, the most-relevant electron operators are $\psi_1\propto e^{-i (k_{F,1}+2k_{F,2})x} e^{-i (\phi_1+2\phi_2)}$ and $\psi_2\propto e^{-i (3k_{F,1}-2k_{F,2})x}e^{-i (3\phi_1-2\phi_2)}$, both with spin up, where $k_{F,1}$ and $k_{F,2}$ are constants. The spin density operators are $S^x=S^y=0$, and $S^z=S^z_f+S^z_b$, where
\begin{align}
S^z_f&=\frac{1}{2}(\psi^{\dagger}_1\psi_1+\psi^{\dagger}_2\psi_2)\propto \partial_x\phi_{\rho}\nonumber\\
S^z_b&=\frac{1}{2}(\psi^{\dagger}_1\psi_2+\textrm{H.c.})\propto e^{i\Delta k x}e^{i2\phi_n}+\textrm{H.c.},
\end{align}
with $\Delta k=-2k_{F,1}+4k_{F,2}$. Using the definition of eigenmodes in Eq.~(\ref{b9}), we find $\mathcal{G}^{x}=\mathcal{G}^{y}=0$, and $\mathcal{G}^{z}=\mathcal{G}^{z}_f+\mathcal{G}^{z}_b$, where
\begin{align}
\mathcal{G}^{z}_f(x,t) &\propto \frac{\tilde{c}_-^2}{(x+v_+t-i \delta)^2}+\frac{\tilde{c}_+^2}{(x-v_-t+i \delta)^2} \nonumber \\
\mathcal{G}^{z}_b(x,t) &\propto\cos(\Delta kx) \Big[\frac{1}{x+v_+t-i  \delta}\Big]^{2\tilde{c}_+^2}\Big[\frac{1}{x-v_-t+i  \delta}\Big]^{2\tilde{c}_-^2}.
\end{align}

\subsubsection{The Pfaffian state at $\nu=5/2$}
The Pfaffian state is spin-polarized and has a chiral edge state. The Lagrangian density for the edge is 
\begin{equation}
L= \frac{2}{4\pi} [\partial_{t}\phi_{1}\partial_{x}\phi_{1}-v_1 (\partial_{x}\phi_{1})^{2}]+ i \lambda (\partial_{t}-v_{\lambda}\partial_{x})\lambda,
\end{equation}
where $\phi_{1}$ is a bosonic charged mode and $\lambda$ is a Majorana fermion. We assume the edge state is left-moving. The most-relevant electron operator is $\psi\propto\lambda e^{-i2\phi_1}$. The spin density operators are $S^x=S^y=0$, and
\begin{align}
S^z=\frac{1}{2}\psi^{\dagger}\psi \propto \partial_x \phi_1 ,
\end{align}
where we have used $\lambda^2=1$. We find $\mathcal{G}^{x}=\mathcal{G}^{y}=0$, and 
\begin{equation}
\mathcal{G}^{z}(x,t)\propto \frac{1}{(x+v_1t-i \delta)^2}.
\end{equation}

\subsubsection{The anti-Pfaffian state at $\nu=5/2$}
The anti-Pfaffian state is the particle-hole dual of the Pfaffian state. The state is spin-polarized. We consider the situation of a clean sample where disorder effect can be neglected and there is translation invariance on the edge. The edge Lagrangian density then takes the form
\begin{align}
\label{b10}
L=  \frac{1}{4\pi}& [2\partial_{t}\phi_{\rho}\partial_{x}\phi_{\rho}-\partial_{t}\phi_{n}\partial_{x}\phi_{n}-2v_{\rho} (\partial_{x}\phi_{\rho})^{2} \nonumber\\
&-v_{n} (\partial_{x}\phi_{n})^{2} -2v_{\rho n}\partial_{x}\phi_{\rho}\partial_{x}\phi_{n}] +i \lambda (\partial_{t}+v_{\lambda}\partial_{x})\lambda,
\end{align}
where $\phi_{\rho}$ is a left-moving charged boson, $\phi_{n}$ is a  right-moving neutral boson and $\lambda$ is a  right-moving Majorana fermion. The edge state is non-chiral. Assuming charge-neutral separation in the edge state, i.e., $v_{\rho}\gg v_{n}\sim v_{\rho n}\sim v_{\lambda}$, we find three most-relevant electron operators: $\psi_1\propto\lambda e^{-i2\phi_{\rho}}$, $\psi_2\propto e^{-i\phi_{n}} e^{-i2\phi_{\rho}}$ and $\psi_3\propto e^{i\phi_{n}} e^{-i2\phi_{\rho}}$. 
The spin density operators are $S^x=S^y=0$, and $S^z=S^z_f+S^z_{b1}+S^z_{b2}$, where
\begin{align}
S^z_f&=\frac{1}{2}(\psi^{\dagger}_1\psi_1+\psi^{\dagger}_2\psi_2+\psi^{\dagger}_3\psi_3)\propto \partial_x\phi_{\rho}\nonumber\\
S^z_{b1}&=\frac{1}{2}(\psi^{\dagger}_1\psi_2+\psi^{\dagger}_1\psi_3+\textrm{H.c.})\propto e^{i\Delta k x}\lambda e^{i\phi_n}+\textrm{H.c.}\nonumber \\
S^z_{b2}&=\frac{1}{2}(\psi^{\dagger}_2\psi_3+\textrm{H.c.})\propto e^{i\Delta k' x}e^{i2\phi_n}+\textrm{H.c.},
\end{align}
with $\Delta k,\Delta k'$ the momentum differences between the edge modes. Upon diagonalizing the edge theory in Eq.~(\ref{b10}), we find
$\mathcal{G}^{x}=\mathcal{G}^{y}=0$, and $\mathcal{G}^{z}=\mathcal{G}^{z}_f+\mathcal{G}^{z}_{b1}+\mathcal{G}^{z}_{b2}$, where
\begin{align}
\label{b16}
\mathcal{G}^{z}_f(x,t) \propto & \frac{\tilde{c}_-^2}{(x+v_+t-i \delta)^2}+\frac{\tilde{c}_+^2}{(x-v_-t+i \delta)^2} \nonumber \\
\mathcal{G}^{z}_{b1}(x,t)\propto & \cos(\Delta kx)\Big[\frac{1}{x+v_+t-i  \delta}\Big]^{\frac{1}{2}\tilde{c}_+^2}\Big[\frac{1}{x-v_-t+i  \delta}\Big]^{\frac{1}{2}\tilde{c}_-^2} \nonumber\\
&\times\frac{1}{x-v_{\lambda}t+i  \delta} \nonumber\\
\mathcal{G}^{z}_{b2}(x,t)\propto &\cos(\Delta k'x) \Big[\frac{1}{x+v_+t-i  \delta}\Big]^{2\tilde{c}_+^2}\Big[\frac{1}{x-v_-t+i  \delta}\Big]^{2\tilde{c}_-^2},
\end{align}
where $\tilde{c}_+(\theta)=\sqrt{2}\sinh{\theta}$ and $\tilde{c}_-(\theta)=\sqrt{2}\cosh{\theta}$. The parameters $v_+\simeq v_{\rho}$, $v_-\simeq v_n$ and $v_+\gg v_-$. Notice that $\mathcal{G}^{z}_f,\mathcal{G}^{z}_{b1}$ dominate over $\mathcal{G}^{z}_{b2}$ at long distances.

\subsubsection{The $SU(2)_2$ state at $\nu=5/2$}

This is a spin-polarized state. The edge Lagrangian density is
\begin{align}
L=  \frac{1}{4\pi} & [2\partial_{t}\phi_{\rho}\partial_{x}\phi_{\rho}+\partial_{t}\phi_{n}\partial_{x}\phi_{n}-2v_{\rho} (\partial_{x}\phi_{\rho})^{2} \nonumber\\
&-v_{n} (\partial_{x}\phi_{n})^{2}] +i \lambda (\partial_{t}-v_{\lambda}\partial_{x})\lambda,
\end{align}
where $\phi_{\rho}$ is a charged boson, $\phi_{n}$ is a neutral boson and $\lambda$ is a Majorana fermion. The edge state is chiral. The most-relevant electron operators and the spin density operators have the same forms as those in the anti-Pfaffian state. However, note that the fields $\phi_{\rho}, \phi_n$ here have different origins from those in Eq.~(\ref{b10}). The correlators are found to be $\mathcal{G}^{x}=\mathcal{G}^{y}=0$, and $\mathcal{G}^{z}=\mathcal{G}^{z}_f+\mathcal{G}^{z}_{b1}+\mathcal{G}^{z}_{b2}$, where
\begin{align}
\mathcal{G}^{z}_f(x,t) &\propto \frac{1}{(x+v_{\rho}t-i \delta)^2} \nonumber \\
\mathcal{G}^{z}_{b1}(x,t)&\propto \cos(\Delta kx)\frac{1}{(x+v_nt-i  \delta)(x+v_{\lambda}t-i  \delta)}  \nonumber\\
\mathcal{G}^{z}_{b2}(x,t)&\propto \cos(\Delta k'x) \frac{1}{(x+v_nt-i  \delta)^4},
\end{align}
with $\Delta k,\Delta k'$ the momentum differences between the edge modes.

\subsection{Time integral}

The QH states we have discussed can be divided into three types. 

Type (i):
The edge state is chiral and contains one bosonic mode. Examples include Laughlin states at $\nu=1/m$ and the Pfaffian state at $\nu=5/2$. The in-plane correlators vanish, while the out-of-plane correlator has the form
\begin{align}
\label{b11}
\mathcal{G}^{\textrm{(i)}}(x,t)= \Big[\frac{1}{\delta+i(t+x/v)}\Big]^{n},
\end{align}
neglecting the proportionality constant and assuming the edge state is left-moving, where $n\geq 2$ is an even integer and $v>0$ is the speed of the edge mode. 

Type (ii):
The edge state is chiral and contains multiple interacting bosonic modes. Examples include the QH state at $\nu=2$ and the 331 state at $\nu=5/2$. The correlators can have the form of Eq.~(\ref{b11}), or 
\begin{align}
\label{b12}
\mathcal{G}^{\textrm{(ii)}}(x,t)=  \Big[\frac{1}{\delta+i(t+x/v_+)}\Big]^{g_+} \Big[\frac{1}{\delta+i(t+x/v_-)}\Big]^{g_-},
\end{align}
neglecting the proportionality constant and the modulating factor, and assuming the edge state is left-moving, where $g_+$ and $g_-$ are non-integers but $g_++g_-$ is an even integer. From previous calculations, we have $0<g_+\ll1$ and $g_->1$. To a good approximation, $v_+$ and $v_-$ can be considered as the speeds of the physical charged mode and neutral mode in the edge state, respectively, so that $v_+\gg v_->0$. We have suppressed the spin-component index for simplicity.

Type (iii):
The edge state is non-chiral. Examples include the QH state at $\nu=2/3$ and the 113 state at $\nu=5/2$. The correlators can have the form of Eq.~(\ref{b11}), or 
\begin{align}
\label{b13}
\mathcal{G}^{\textrm{(iii)}}(x,t)=  \Big[\frac{1}{\delta+i(t+x/v_+)}\Big]^{g_+} \Big[\frac{1}{\delta+i(t-x/v_-)}\Big]^{g_-},
\end{align}
neglecting the proportionality constant and the modulating factor, where $g_+$, $g_-$, and $g_++g_-$ are all non-integers. We have $0<g_+\ll1$ and $g_->1$. The parameters $v_+$ and $v_-$ can again be considered as the speeds of the physical charged mode and neutral mode, respectively, so that $v_+\gg v_->0$.

In writing Eqs.~(\ref{b11})$-$(\ref{b13}), we have assumed that there are only two distinct velocities in the system: The charged-mode velocity and the neutral-mode velocity. In particular, for the anti-Pfaffian state we make the  approximation that the Majorana fermion and the neutral boson propagate at the same speed. For the $SU(2)_2$ state, there is no need for such an approximation and the correlators take the forms of either Eq.~(\ref{b11}) or Eq.~(\ref{b12}).

In the following, we evaluate $I=\int^{\infty}_0dt e^{-\eta t} \textrm{Im} \mathcal{G}^a(x,t)$, where $\eta=0^+$ and $a=$(i), (ii), (iii).

\subsubsection{Type (i) }
For type (i) edge states, 
\begin{align}
I&= \frac{1}{2i}\{\int^{\infty}_0dt\  e^{-\eta t} \Big[\frac{1}{\delta+i(t+x/v)}\Big]^{n}-\textrm{c.c.} \}\nonumber\\
&\equiv I_1-I_2.
\end{align}
The integrand of $I_1$ has an $n$th-order pole at $t_1=-x/v+i\delta$, while the integrand of $I_2$ has an $n$th-order pole at $t_2=-x/v-i\delta$. By Residue theorem, 
\begin{equation}
 \int_0^{\infty}dt\ \frac{1}{(t-t_k)^n}=-\textrm{Res}(\frac{ \ln t}{(t-t_k)^n}; t_k),
 \end{equation}
where $k=1,2$. This gives 
\begin{align}
I_1=I_2=\frac{1}{2i}\frac{(-1)^{n/2}}{n-1}\Big(\frac{x}{v}\Big)^{1-n} ,
\end{align}
so that  $I=0$. This suggests that the spin susceptibility in type (i) edge states vanishes to the lowest order (i.e., considering only the most-relevant operators).

\subsubsection{Type (ii) }
For type (ii) edge states, 
\begin{equation}
I =\int^{\infty}_0dt\ e^{-\eta t} \textrm{Im}\Big[\frac{1}{\delta+i(t+x/v_+)}\Big]^{g_+} \Big[\frac{1}{\delta+i(t+x/v_-)}\Big]^{g_-}.
\end{equation}
The integrand has two branch points $-x/v_++i\delta$ and $-x/v_-+i\delta$. Choosing the branch cut appropriately,
\begin{align}
& \textrm{Im}\Big[\frac{1}{\delta+i(t+x/v_+)}\Big]^{g_+} \Big[\frac{1}{\delta+i(t+x/v_-)}\Big]^{g_-} \nonumber \\
&=  \textrm{Im}\{ e^{-i \frac{\pi}{2}g_+ \textrm{sgn}(t+\frac{x}{v_+})}e^{-i \frac{\pi}{2}g_- \textrm{sgn}(t+\frac{x}{v_-})}\} |\mathcal{G}^{\textrm{(ii)}}(x,t)|\nonumber \\
&=\Theta(t+\frac{x}{v_+})\Theta(-t-\frac{x}{v_-}) \sin[\frac{\pi}{2}(g_--g_+)]|\mathcal{G}^{\textrm{(ii)}}(x,t)| ,
\end{align}
where $\Theta(x)$ is the Heaviside step function, $\textrm{sgn}(x)$ is the signum function, and 
\begin{equation}
|\mathcal{G}^{\textrm{(ii)}}(x,t)|=\Big|t+\frac{x}{v_+}\Big|^{-g_+}\Big|t+\frac{x}{v_-}\Big|^{-g_-} .
\end{equation}
 We have used the fact that $g_++g_- $ is an even integer, so that $\textrm{Im} \{ e^{-i \frac{\pi}{2}(g_++g_-)}\}=0$. Notice also that $I=0$ if we set $v_+=v_-$, which is consistent with the previous result for type (i) edge states. In our scenario, $v_+\gg v_->0$. The integral is nonzero only when $x<0$. Explicitly,  
\begin{align}
\label{b14}
I=& \Theta(-x) \sin[\frac{\pi}{2}(g_--g_+)] \int_{-x/v_+}^{-x/v_-}dt\ e^{-\eta t}|\mathcal{G}^{\textrm{(ii)}}(x,t)| \nonumber\\ 
 = &\Theta(-x) \sin[\frac{\pi}{2}(g_--g_+)] (\frac{1}{v_+}-\frac{1}{v_-})^{-g}  \nonumber \\ &\times B(1-g_+,1-g_-) |x|^{-g} ,
\end{align}
where $B(x,y)$ is the Euler beta function and we define $g=g_++g_--1$. 

The above calculation applies to left-moving edge states. For right-moving edge states, one replaces $\Theta(- x)$ with $\Theta(x)$, and sends $v_+,v_-\rightarrow -v_+,-v_-$ in Eq.~(\ref{b14}). The exponent $g$ determines the scaling of the spin susceptibility with distance, and may take different values $g^{\alpha}$ for different spin components $\alpha=x,y,z$. For type (ii) edge states, $g^{\alpha}$ are integral invariants depending on the topological order of the bulk QH liquid. For instance, $g^x=g^y=1$ for the QH state at $\nu=2$.

\subsubsection{Type (iii) }
For type (iii) edge states, 
\begin{equation}
I= \int^{\infty}_0dt\ e^{-\eta t} \textrm{Im}\Big[\frac{1}{\delta+i(t+x/v_+)}\Big]^{g_+} \Big[\frac{1}{\delta+i(t-x/v_-)}\Big]^{g_-}.
\end{equation}
We have
\begin{align}
 \textrm{Im}&\Big[\frac{1}{\delta+i(t+x/v_+)}\Big]^{g_+} \Big[\frac{1}{\delta+i(t-x/v_-)}\Big]^{g_-} \nonumber \\
= & \textrm{Im}\{ e^{-i \frac{\pi}{2}g_+ \textrm{sgn}(t+\frac{x}{v_+})}e^{-i \frac{\pi}{2}g_- \textrm{sgn}(t-\frac{x}{v_-})}\} |\mathcal{G}^{\textrm{(iii)}}(x,t)| \nonumber \\
=&\{ [\Theta(-t-\frac{x}{v_+}) -\Theta(-t+\frac{x}{v_-}) ]\sin[\frac{\pi}{2}(g_+-g_-)] \nonumber \\
&-\Theta(t+\frac{x}{v_+})\Theta(t-\frac{x}{v_-}) \sin[\frac{\pi}{2}(g+1)] \}  |\mathcal{G}^{\textrm{(iii)}}(x,t)|,
\end{align}
where 
\begin{equation}
|\mathcal{G}^{\textrm{(iii)}}(x,t)|=\Big|t+\frac{x}{v_+}\Big|^{-g_+}\Big|t-\frac{x}{v_-}\Big|^{-g_-} .
\end{equation}
The integral is nonzero for both $x>0$ and $x<0$. We find $I=\Theta(x)I_>+\Theta(-x)I_<$, where 
\begin{align}
\label{b15}
I_>  =&  \sin[\frac{\pi}{2}(g_--g_+)] \int_{0}^{x/v_-}dt\ e^{-\eta t} |\mathcal{G}^{\textrm{(iii)}}(x,t)| \nonumber\\
&-\sin[\frac{\pi}{2}(g+1)] \int_{x/v_-}^{\infty}dt\ e^{-\eta t} |\mathcal{G}^{\textrm{(iii)}}(x,t)| \nonumber\\ 
 =& \{\sin[\frac{\pi}{2}(g_--g_+)] \frac{v_+^{g_+}v_-^{g_--1}}{1-g_-}  F(1,g_+;2-g_-;-\frac{v_+}{v_-}) \nonumber \\
&-\sin[\frac{\pi}{2}(g+1)]   (\frac{1}{v_+}+\frac{1}{v_-})^{-g}B(g,1-g_-) \} |x|^{-g},  
\end{align}
and 
\begin{align}
\label{b15b}
I_<  =&  \sin[\frac{\pi}{2}(g_+-g_-)] \int_{0}^{-x/v_+}dt\ e^{-\eta t} |\mathcal{G}^{\textrm{(iii)}}(x,t)|\nonumber\\
& -\sin[\frac{\pi}{2}(g+1)] \int_{-x/v_+}^{\infty}dt\  e^{-\eta t} |\mathcal{G}^{\textrm{(iii)}}(x,t)|\nonumber\\ 
 =& \{\sin[\frac{\pi}{2}(g_+-g_-)] \frac{v_+^{g_+-1}v_-^{g_-}}{1-g_+}  F(1,g_-;2-g_+;-\frac{v_-}{v_+})  \nonumber \\
&-\sin[\frac{\pi}{2}(g+1)]   (\frac{1}{v_+}+\frac{1}{v_-})^{-g}B(g,1-g_+)\} |x|^{-g}   ,
\end{align}
where $F(a,b;c;x)$ is the hypergeometric function. Notice that $I_>$ and $I_<$ are related by the exchange of parameters
\begin{equation}
g_+\leftrightarrow g_- \mspace{9.0mu} \textrm{and} \mspace{9.0mu}
v_+\leftrightarrow v_-,
\end{equation}
which technically reverts the chirality of all the edge modes, as seen from Eq.~(\ref{b13}). For type (iii) edge states, $g$ (i.e., $g^{\alpha}$, where $\alpha=x,y,z$) takes non-integer values. Let us write $g^{\alpha}=g^{\alpha}_0+\delta g^{\alpha}$, where $g^{\alpha}_0$ is the integer part of $g^{\alpha}$. We find $\delta g^{\alpha}\ll  g^{\alpha}_0$ for all the type (iii) edge states being discussed. For instance, $\delta g^x=\delta g^y=4\sinh^2\theta$, where $\theta\ll 1$, while $g^x_0=g^y_0=1$ in the spin-unpolarized QH state at $\nu=2/3$.

\subsection{Full expression of spin susceptibility}

Substituting the above results in Eq.~(\ref{b0}), we obtain the spin susceptibility in QH edge states, 
\begin{align}
\chi^{\alpha\alpha}(x)=\frac{\cos(\Delta kx)}{4\pi^2}l^2a^{g^{\alpha}-1}v_+^{-g_+^{\alpha}}v_-^{-g_-^{\alpha}}\times I,
\end{align}
where we have restored the spin-component index and the proportionality constant.
The short-distance cut-off $a$ can be taken as the lattice constant of the host material of the QH system. For left-moving type (ii) edge states, $I$ is given by Eq.~(\ref{b14}). We have $\chi^{\alpha\alpha}(x)=\cos(\Delta kx)|x|^{-g^{\alpha}}\Theta(-x)C^{\alpha} (\boldsymbol{g}^{\alpha}, \boldsymbol{v})$, where $\boldsymbol{g}^{\alpha}=(g_+^{\alpha},g_-^{\alpha})$, $\boldsymbol{v}=(v_+,v_-)$, and
\begin{align}
\label{b17}
C^{\alpha} (\boldsymbol{g}^{\alpha}, \boldsymbol{v})=&\frac{l^2a^{g^{\alpha}-1}}{4\pi^2}\sin[\frac{\pi}{2}(g_-^{\alpha}-g_+^{\alpha})]\frac{v_+^{g_-^{\alpha}-1}v_-^{g_+^{\alpha}-1}}{(v_--v_+)^{g^{\alpha}}}\nonumber \\ &\times B(1-g_+^{\alpha},1-g_-^{\alpha}).
\end{align}
For type (iii) edge states, $I$ is given by Eqs.~(\ref{b15})(\ref{b15b}). We have $\chi^{\alpha\alpha}(x)=\cos(\Delta kx)|x|^{-g^{\alpha}}\{\Theta(x) C^{\alpha} _>(\boldsymbol{g}^{\alpha}, \boldsymbol{v})  +\Theta(-x) C^{\alpha} _<(\boldsymbol{g}^{\alpha}, \boldsymbol{v})  \}$, where
\begin{widetext}
\begin{align}
\label{b18}
C^{\alpha}_>(\boldsymbol{g}^{\alpha}, \boldsymbol{v})=&\frac{l^2a^{g^{\alpha}-1}}{4\pi^2} \{\sin[\frac{\pi}{2}(g_-^{\alpha}-g_+^{\alpha})] \frac{v_-^{-1}}{1-g_-^{\alpha}}  F(1,g_+^{\alpha};2-g_-^{\alpha};-\frac{v_+}{v_-}) 
-\sin[\frac{\pi}{2}(g^{\alpha}+1)]  \frac{v_+^{g_-^{\alpha}-1}v_-^{g_+^{\alpha}-1}}{(v_++v_-)^{g^{\alpha}}}B(g^{\alpha},1-g_-^{\alpha}) \}\nonumber \\
C^{\alpha}_<(\boldsymbol{g}^{\alpha}, \boldsymbol{v})=&\frac{l^2a^{g^{\alpha}-1}}{4\pi^2} \{\sin[\frac{\pi}{2}(g_+^{\alpha}-g_-^{\alpha})] \frac{v_+^{-1}}{1-g_+^{\alpha}}  F(1,g_-^{\alpha};2-g_+^{\alpha};-\frac{v_-}{v_+}) -\sin[\frac{\pi}{2}(g^{\alpha}+1)]  \frac{v_+^{g_-^{\alpha}-1}v_-^{g_+^{\alpha}-1}}{(v_++v_-)^{g^{\alpha}}} B(g^{\alpha},1-g_+^{\alpha}) \}.
\end{align}
\end{widetext}
We see that $C^{\alpha}_>(\boldsymbol{g}^{\alpha}, \boldsymbol{v})$ and $C^{\alpha}_<(\boldsymbol{g}^{\alpha}, \boldsymbol{v})$ are related by the exchange of arguments: $g_+^{\alpha}\leftrightarrow g_-^{\alpha}$ and $ v_+\leftrightarrow v_-$.

Eqs.~(\ref{b17}) and (\ref{b18}) show that the RKKY interaction is ferromagnetic at short distances.  

To estimate the strength of the RKKY interaction, we extract the dimensional part $[\chi^{\alpha\alpha}(x)]$ of the spin susceptibility. For type (ii) edge states, 
\begin{align}
[\chi^{\alpha\alpha}(x)]=l^2a^{g^{\alpha}-1}\frac{v_+^{g_-^{\alpha}-1}v_-^{g_+^{\alpha}-1}}{(v_+-v_-)^{g^{\alpha}}}|x|^{-g^{\alpha}}.
\end{align}
For type (iii) edge states, there are multiple terms in $\chi^{\alpha\alpha}(x)$, with 
\begin{align}
[\chi^{\alpha\alpha}(x)]=&l^2a^{g^{\alpha}-1}\frac{v_+^{g_-^{\alpha}-1}v_-^{g_+^{\alpha}-1}}{(v_++v_-)^{g^{\alpha}}}|x|^{-g^{\alpha}}; \nonumber\\
&l^2a^{g^{\alpha}-1}v_-^{-1}|x|^{-g^{\alpha}};l^2a^{g^{\alpha}-1}v_+^{-1}|x|^{-g^{\alpha}}.
\end{align}
Using $0<g_+^{\alpha}\ll1$ and $v_+\gg v_-$, we find 
\begin{align}
[\chi^{\alpha\alpha}(x)]\simeq l^2a^{g^{\alpha}-1}v_-^{-1}|x|^{-g^{\alpha}},
\end{align}
for both type (ii) and type (iii) edge states.

\section{Exchange}

Here we estimate the strength of the exchange interaction between the QD electron and the electrons in the edge modes.
The textbook formula gives the exchange integral as
\begin{equation}
J = C \int d{\bf r}_1  d{\bf r}_2\ \Psi_1^*\left({\bf r}_1\right) \Psi_2^*\left({\bf r}_2\right) \frac{1}{|{\bf r}_1-{\bf r}_2|} \Psi_1\left({\bf r}_2\right) \Psi_2\left({\bf r}_1\right),
\label{eq:Jnonlocal}
\end{equation}
for two particles in single particle orbitals $\Psi_1$, $\Psi_2$, interacting through an unscreened Coulomb interaction parametrized by $C=\frac{e^2}{4 \pi \epsilon_0 \epsilon_r}$, where $e$ is the elementary charge, $\epsilon_0$ the vacuum permittivity, and $\epsilon_r$ the relative permittivity of the medium.

Providing a microscopic theory of the exchange for our case is well beyond the scope of this article. Instead we are interested only in the interaction strength scale. To get a rough estimate, let us assume that the exchange interaction is local
\begin{equation}
J = \beta \int  d{\bf r}_1 d{\bf r}_2  \Psi_1^*\left({\bf r}_1\right) \Psi_2^*\left({\bf r}_2\right) \delta\left( {\bf r}_1-{\bf r}_2\right) \Psi_1\left({\bf r}_2\right) \Psi_2\left({\bf r}_1\right),
\end{equation}
which transforms the equation into a density-density interaction
\begin{equation}
J = \beta \int  d{\bf r} \ \rho_1\left({\bf r}\right) \rho_2\left({\bf r}\right).
\label{eq:Jlocal}
\end{equation}
One can explicitly evaluate Eq.~\eqref{eq:Jnonlocal} for a tunnel coupled double dot modelled by a 2D harmonic confinement, \cite{dimitrije12} and then compare to the result given by Eq.~\eqref{eq:Jlocal}. The calculated energies scale the same with the interdot distance and the overall prefactors are related by $\beta=C l$, with $l$ the confinement length of the dot potential. We further guide ourselves by experiments, which measured the exchange energy in few electron QDs made in 2DEG in GaAs. The maximal scale $C/l$, which evaluates to $\simeq3$~meV for typical GaAs parameters $\epsilon_r=12.9$ and $l=30$~nm, is indeed approached in a single dot where the densities overlap in Eq.~\eqref{eq:Jlocal} is of order 1 in dimensionless units ($l^{-2}$). A suppression of the interdot tunneling (by increasing the interdot distance) leads to a decreasing exchange, which reaches $J_{DD} \simeq 0.1-0.01$~meV in a tunnel coupled double dot. Assuming that an analogous suppression will result from the tunnel coupling of our dot coupled to the edge finally gives $J_{DD}$ as an order of magnitude estimate for 
$\Gamma $, which we used in the main text as the coupling constant between an electron spin in a QD and a quasi-1D spin density of the edge.



\end{document}